%% file: manuscript.tex
\documentclass[fleqn,usenatbib]{mnras}

\include{newcommands}

\usepackage{newtxtext,newtxmath,lineno}

\usepackage[T1]{fontenc}

\usepackage{graphicx}	
\usepackage{amsmath}	
\usepackage{multicol}        
\usepackage{bm}		
\usepackage{pdflscape}	
\usepackage{longtable}
\usepackage{xtab}

\newcommand{\tauconv}{$\tau_{\rm c}$}
\newcommand{\prot}{$P_{\rm rot}$}

\usepackage{ae,aecompl}
\usepackage{breqn}

\title[Cool Exoplanet Host Star Ages]{The TIME Table: Rotation and Ages of Cool Exoplanet Host Stars}

\author[E. Gaidos et al.]{Eric Gaidos\thanks{Contact e-mail: gaidos@hawaii.edu}$^{1,2,3}$,
Zachary Claytor$^{4,5}$,
Ryan Dungee$^{6}$,
Aleezah Ali$^{4}$,
Gregory A. Feiden$^{7}$
\\
$^{1}$Department of Earth Sciences, University of Hawai'i at M\={a}noa, Honolulu, HI  96822, USA\\
$^{2}$Institute for Astronomy, University of Vienna, 1180 Wien, Austria\\
$^{3}$Institute for Particle Physics \& Astrophysics, ETH Z\"{u}rich, 8093 Z\"{u}rich, Switzerland\\
$^{4}$Institute for Astronomy, University of Hawai'i at M\={a}noa, Honolulu, HI 96822 USA\\
$^{5}$Department of Astronomy, University of Florida, 211 Bryant Space Science Center, Gainesville, FL 32611 USA\\
$^{6}$Institute for Astronomy, University of Hawai'i at Hilo, Hilo, HI 96720 USA\\
$^{7}$Department of Physics and Astronomy, University of North Georgia, Dahlonega, GA 30597 USA
}
\date{Submitted, accepted}

\pubyear{2022}

\begin{document}
\label{firstpage}
\pagerange{\pageref{firstpage}--\pageref{lastpage}}
\maketitle

\begin{abstract} 
Age is a stellar parameter that is both fundamental and difficult to determine.  Among middle-aged M dwarfs, the most prolific hosts of close-in and detectable exoplanets, gyrochronology is the most promising method to assign ages, but requires calibration by rotation-temperature sequences (gyrochrones) in clusters of known ages.  We curated a catalog of 249 late K- and M-type (\teff=3200-4200K) exoplanet host stars with established rotation periods, and applied empirical, temperature-dependent rotation-age relations based on relevant published gyrochrones, including one derived from observations of the 4 Gyr-old open cluster M67.  We estimated ages for 227 of these stars, and upper limits for 8 others, excluding 14 which are too rapidly rotating or are otherwise outside the valid parameter range of our gyrochronology.  We estimated uncertainties based on observed scatter in rotation periods in young clusters, error in the gyrochrones, and uncertainties in temperature and non-solar metallicity.  For those stars with measured metallicities, we provide but do not incorporate a correction for the effects of deviation from solar-metallicity.  The age distribution of our sample declines to near zero at 10~Gyr, the age of the Galactic disk, with the handful of outliers explainable by large uncertainties.  Continued addition or extension of cluster rotation sequences to more thoroughly calibrate the gyrochronology in time and temperature space, more precise and robust measurement of rotation periods, and more accurate stellar parameter measurements will enable continued improvements in the age estimates of these important exoplanet host stars.
\end{abstract}





\section{Introduction}

Over the past three decades, thousands of planets have been discovered around other stars.   Exoplanet surveys have revealed that M-type dwarfs, the least massive but most numerous stars, host more planets on close-orbits than their solar-mass counterparts \citep{Mulders2015,Hardegree-Ullman2019,Hsu2019}.  These include Earth-size, rocky planets that orbit within the compact habitable zone of these intrinsically faint stars, and which are more feasible to study, e.g. with \jwst, because of the host stars' lower mass, radius, and luminosity.    

Precise characterization of the host star (e.g., radius, mass, luminosity, metallicity) is essential to obtain properties of its planets, but is challenging for very low-mass stars for which methods tuned to the Sun do not apply.  For this reason, empirical approaches have proven useful for M dwarfs, enabled by the advent of the \gaia\ astrometry mission, space- and ground-based photometric surveys, and advances and expansion in spectroscopic instrumentation.  For example, interferometry can directly measure the angular radii of very nearby M dwarfs; pairing with trigonometric parallaxes yields physical radii.  Combined with a bolometric luminosity from a flux-calibrated spectral energy distribution (SED), this allows the effective temperature \teff\ to be derived using the Stefan-Boltzmann law  \citep{Boyajian2012}.  Spectra of stars with  a range of established \teff\ can then serve as templates to estimate the \teff\ of more distant stars using spectra and, combining with SEDs, their radii \citep{Mann2015}.  Metallicities can be calibrated using binaries where the solar-type companion has an established metallicity (relative to the Sun) \citep[e.g.,][]{Mann2013c,Mann2014,Montes2019,Souto2020}.

Age is a fundamental property of planets but is difficult to accurately estimate for most systems \citep{Christensen-Dalsgaard2018}.  Planets and their atmospheres are expected to evolve under the influence of their host star and their own internal thermodynamics and compositional change \citep{Kite2009,Lammer2013}.  For temperate, Earth-like planets, changes in atmospheric composition will be the backdrop against which biosignatures will be searched for.  Observations of disk lifetimes, planet formation theory, and isotope-ages of Solar System bodies indicate that the age of a star should be no more than a few tens of Myr older than that of its planets \citep{Helled2021}, but ages of most host stars remain poorly constrained.  Planets are difficult to detect around young stars \citep[e.g.,][]{Miyakawa2022} and most planet hosts are older and no longer members of (relatively) well-dated clusters and young-moving groups.

Ages of isolated field stars have been estimated by (1) comparison of stellar parameters to stellar models (e.g., in a color-magnitude diagram); (2) asteroseismic measurement of increasing density due to the conversion of H into He and heavy elements in stellar interiors; (3) the abundance of lithium, which is destroyed in stellar interiors; (4) metallicity and the overall age-metallicity relation of the Galactic disk; (5) increase in the peculiar motion of stars with time with respect to the overall orbital motion of the Galactic disk as a result of perturbations from molecular clouds and other stars; and (6) rotation and rotation-driven magnetic activity that decline as angular momentum (AM) is lost through a magnetized wind.

But M dwarfs are resistant to most age-dating techniques:  They evolve imperceptibly on the main sequence \citep[MS;][]{Adams2005} and the pulsations that are the grist of asteroseismology are below current detection thresholds \citep[][and regardless the stars' densities do not change]{Rodriguez-Lopez2019} .  M dwarfs consume their lithium within $\sim$50 million years \citep{Binks2014} and thus this proxy cannot be used at later ages, and metallicity and peculiar motions are only meaningful in a statistical sense for stellar populations, not individual stars; the age-metallicity relation appears flat for disk stars \citep{Rebassa-Mansergas2021} and age-abundance ratio relations do not appear to apply universally \citep{Casali2020}.

This leaves gyrochronology, the application of relations between age and rotation (and its proxies) brought about by stellar spin-down, as a viable method to age-date main-sequence M dwarfs in the field \citep{Barnes2007}.  Such stars are potentially well-suited for this approach because they have a smaller or no inner radiative core which can rotate quasi-independently of the convective envelope, and instead could have behavior similar to simple ``solid-body" rotation.    However, spin-down driven by magnetic activity scales with the Rossby number $Ro \equiv P_{\rm rot}/\tau_c$, where $\tau_c$ is the local turnover time in the convective envelope, and the longer $\tau_c$ of M dwarfs compared to solar-type stars means their rotational evolution will be distinct.  Using co-eval bainries, \citet{Otani2022} tested several color-dependent spin-down models for internal consistency, and derived \emph{internal} errors (i.e., only those arising from errors in \prot\ and color) of 5-10\% for relatively young early M-type stars. 

Pioneering ground-based observations of open clusters of co-eval stars, followed by the revolutionary wide-field surveys of the \kepler\ and \tess\ space telescopes, document the formation of tight sequences in rotation vs. color or \teff\ diagrams that extend to cooler temperatures and longer $\tau_c$ with time \citep{Gallet2015,Curtis2020}.  The formation of the sequence among solar-mass stars is thought to be the result of a change in the braking law (i.e., the rotation-rate dependence of the torque) as stars transition from a ``saturated" (less rotation rate-dependent) phase to an ``unsaturated" (more rotation rate-dependent) phase of stellar activity at a critical value of $Ro$ \citep{Matt2012,Curtis2019}.  After a star enters this regime the strong rotation rate dependence effectively erases the effect of initial conditions, allowing gyrochronologic relations to be applied.    

Among cooler dwarf stars the situation is more complex. The outer convective envelope---the part of the star that is both observable and feels the decelerating torque from the magnetized wind---is more substantial, and the coupling timescale between the the envelopes and the radiative core is longer.  Among K dwarfs, this can lead to early differential rotation, with the core spinning faster than the envelope, and later ``stalling" as the core transfers AM to the envelope and the observed spin-down temporarily slows or halts \citep{Denissenkov2010,Curtis2019b,Curtis2020}.  Stalling could contribute to the formation of a rotational sequence, but also delays the epoch at which gyrochronology is useful.  Among yet cooler M dwarfs the radiative core is smaller or absent, the coupling timescale is expected to get longer, and the magnitude of the stalling could diminish/disappear \citep{Lu2022}.

The mechanism responsible for core-envelope coupling has not been established, nor has a theoretical model that is quantitatively consistent with the observations and has predictive power over a range of \teff\ been constructed.  Moreover, it is not certain if braking laws developed for solar-type stars apply to M dwarfs.  For these reasons, observations of stars in clusters of known ages are imperative for identifying rotational sequences (vs. \teff), constructing empirical rotation-age relations, and calibrating successful models.  But M dwarfs are intrinsically faint and their rotation evolves more slowly, and thus deeper observations of older (and statistically more distant) clusters are required.  M dwarfs in these clusters are beyond the range of current space telescopes both in terms of signal (limited by the modest telescope aperture) and spatial resolution (limited by the large pixel size).   \ktwo\ monitoring of the oldest nearby cluster \citep[Ruprecht 147, $\approx$2.7 Gyr;][]{Curtis2013} captured rotation periodsfor only a handful of M dwarfs near the K-M boundary.  Similar observations of the nearest old cluster \citep[M67, $\approx$4 Gyr][]{Richer1998,Vandenberg2004,Schiavon2004,Sarajedini2009} failed to yield useful results \citep{Esselstein2018}.

Ground-based observatories can go deeper and with high resolution. \citet{Dungee2022} carried out a Sloan $i$-band monitoring campaign of late K and early M dwarf members of M67 with the MegaCAM wide-field camera at the prime focus of the 3.6-m Canada France Hawaii Telescope (CFHT) \citep{Boulade1998}, obtaining 294 rotation periods and identifying a rotational sequence that ranged from $\approx$25 days at 4200K to 125 days at 3200K.  \citet{Dungee2022} found that the ``warm" end of the M67 sequence could be explained by the overlapping ``cool" end of the Ruprecht 147 sequence identified by \citet{Curtis2020}, plus Skumanich-like power-law spin-down $P \propto t^{n}$ \cite{Skumanich1972} with an index $n= 0.62$.  This indicates that (a) the rotation sequence of M dwarfs extends close to the fully convective boundary (near \teff=3200K) by no later than 4 Gyr; (b) spin-down among middle-aged M dwarfs seems to obey a relatively simple braking law.  Both of these findings bode well for the gyrochronology of very cool dwarfs.

In this work, we curate a catalog of rotation periods of late K and early M-type dwarfs known to host validated or confirmed planets\footnote{A ``validated" planet is one for which known false positive scenarios are highly unlikely; a ``confirmed" planet has been detected by a second, independent method.}, and apply empirical rotation-age relations based on the M67 gyrochrone of \citet{Dungee2022} and previously published gyrochrones \citep{Curtis2020} to estimate ages.  The rotation periods of many host stars have been established either using the same space-based photometry (i.e., \kepler, \ktwo, \tess) used to identify their transiting planets, or by data obtained from the ground- or space as part of the validation/confirmation of candidate planets.  There are also collections of rotation periods of field stars (including planet hosts) based on data from \kepler\ \citep{Santos2019,Santos2021}, \ktwo\ \citep{Reinhold2020}, \tess\ \citep{Canto-Martins2020}, and ground-based surveys  \citep{Oelkers2018,Newton2018,Christy2022}.  We also identify additional candidate rotational signatures directly in the photometric datasets.  We emphasize that some rotation periods are tentative and that very cool dwarf gyrochronology is a work in progress and makes assumptions which will be borne out or refuted by future observations.      

\section{Sources of Rotation Periods}
\label{sec:catalog}

We identified all host stars of validated or confirmed exoplanets with \teff\ of 3200--4200\,K in the NASA Exoplanet Archive as of August 2022.  This included 112 \kepler\ host stars or \kepler\ Objects of Interest (KOIs) having rotation periods \prot\ in \citet{Santos2019}.  From the list of the non-\kepler\ host stars we removed evolved (giant), T Tauri, and pre-main sequence (PMS) stars, as rotation of these stars obviously does not follow the gyrochronology of dwarfs, as well as members of star-forming regions and young moving groups that have ages estimated by other techniques, leaving 215 non-\kepler\ stars.  

Rotation periods for these stars were obtained from the literature; these were determined using ground- or space-based photometry, e.g. by WASP \citep{Pollacco2006}, MEarth \citep{Berta2012}, ASAS-SN \citep{Shappee2014}, \ktwo \citep{Howell2014}, and \tess\ \citep{Ricker2014} and/or time-series spectroscopy of indicators of active regions and magnetic fields, notably by the HARPS \citep{Pepe2000} and CARMENES Doppler RV surveys for exoplanets \citep{Reiners2018}.

We revisited the \ktwo\ data by matching all stars against the EPIC catalog \citep{Huber2016} and downloaded all Pre-search Data Conditioning Simple Aperture Photometry (PDCSAP) lightcurves from the MAST archive (including many \ktwo-detected transiting exoplanet host stars).  The lightcurves were further de-trended with a best-fit second-order polynomial before a Lomb-Scargle analysis \citep{Scargle1982} to search for signals with periods of 0.4-40 days, an interval chosen to avoid the pervasive 6-hr thruster firing signal and for the typical lightcurve to span at least two periods.  115 lightcurves of 95 EPIC stars contained peaks exceeding a $p=0.001$ false-positive level.  Of these 20 were not previously published, and in one other case (K2-345) we replaced the \citet{Reinhold2020} value as being obviously erroneous.  We did not revise other \citet{Reinhold2020} values.  In 11 cases (K2-5, 14, 83, 124, 125, 129, 151, 288B, 315, 322, 377) we judged that the peak was an upper harmonic and doubled the period and its error based on inspection of the lightcurve.  The de-trended lightcurves and periodograms are shown in Figs. \ref{fig:ktwo1}-\ref{fig:ktwo4}.   

We retrieved lightcurves from the Zwicky Transient Facility \citep[ZTF,][]{Masci2019} using the {\tt Python} wrapper of the InfraRed Science Archive (IRSA) API query \citep{Rigault2018} and a matching criteria of 5".  We obtained 319 ZTF lightcurves for 102 stars; each star has up to three ($gri$) lightcurves and sometimes the 5" search cone contained more than one star.  We constructed Lomb-Scargle periodograms of the 145 lightcurves with at least 100 points (the maximum was 1092) and identified 25 signals among 22 stars with peaks with a false alarm probability $< 0.01$.  (We were less stringent than for \ktwo\ because of the availability of data in multiple filters for comparison).   For three stars with two periodic signals (in different filters), none had matching periods.  Seven signals were rejected based on similarity to the lunar synodic period (29.5 day) or its seasonal alias (27.3 and 32.1 days).  Marginal, long term ($>100$-day) signals seen in the lightcurves of K2-123 and K2-124 were rejected based on detection of shorter rotation periods by \ktwo.  ZTF periods for two other systems (TOIs-2136 and 3174) were already reported in the literature.  We assigned tentative periods to four other stars based on these data; none of these periods are close to the 29.5-day lunar synodic period or its annual alias, or harmonics of 1-day (Fig. \ref{fig:ztf}).  All other periodograms contained no clearly visible peaks or a forest of peaks of roughly equal (in)significance.  We also retrieved $g$- and $V$-band lightcurves from the All-Sky Automated Search for Super-Novae \citep[ASAS-SN][]{Shappee2014,Kochanek2017} on which we performed a similar analysis.  We identified 16 and 15 host stars with significant ($p<0.01$) periodic signals in the $g$- and $V$-band data, respectively.  Among these are four stars that have matching $g$- and $V$-band periods that are not at the lunar synodic period, its aliases, or 1-day harmonics.  The $g$-band photometry for these is shown in Fig. \ref{fig:asas-sn}.  In the case of GJ 486, the detected signals at 13.7 days differ markedly from the published \prot\ of $49.9\pm5.5$ days \citep{Caballero2022} (which we retain).  

We included \prot\ values determined from time-series measurements of spectroscopic indicators of stellar activity such as \ion{Ca}{II} HK and \halpha, but we excluded estimates based on \emph{correlations} with indicators of stellar activity, e.g., the overall level of \ion{Ca}{II} HK emission \citep[e.g.,][]{Astudillo-Defru2017}.

\section{Stellar Parameters}
\label{sec:parameters}

We retrieved values for \teff\ and metallicities ([Fe/H]) of KOIs from the literature via the NASA Exoplanet Archive tables \citep{Thompson2018,Coughlin2016,Mullally2015,Rowe2015,Burke2014,Batalha2013}. For each star, we used the most recent KOI table available. However, the present characterization of cool exoplanet host stars is markedly heterogeneous, with many stars lacking published metallicities, and multiple values for the same star can differ by much more than the published uncertainties. For those stars without effective temperature solutions from the NASA Exoplanet Archive, we used their absolute $K_s$ magnitude to calculate an effective temperature with the \cite{Mann2015} empirical relations. 

We identified stars in multiple systems using the literature, as well as stars with \gaia\ astrometric renormalized unit weight error (RUWE) values $>$1.4 indicative of unresolved multiplicity \citep{Belokurov2020}. We flagged but did not exclude these systems, cautioning that binaries that are unresolved in the data used to obtain rotational signals could be assigned incorrect rotation periods, and a single-star gyrochronology is expected to be more erroneous or fail in sufficiently close binaries (see Section \ref{sec:summary}).

We searched for additional, unpublished stellar companions resolved by \gaia\  (separations $\gtrsim$1\arcsec) and identifiable based on similar parallaxes and proper motions in the DR3 catalog \citep{Gaia2016,Gaia2022}. This was performed by calculating a Bayesian probability that a candidate companion's astrometry is the same as a given star, relative to the probability that this occurs in a ``background" population.  We identified five stars with FAP $<0.01$, but all are previously known binaries. 

We searched the \emph{Hipparcos}-\emph{Gaia} (EDR3) Catalog of Accelerations \citep{Brandt2021} and found 5 matches, a paucity that reflects the magnitude limit of the \emph{Hipparcos} catalog.  Of these, only two have $\chi^2$ fits approaching but not reaching a formal 0.3\% false-alarm probability threshold of 11.8: HD 238090 ($\chi^2=9.1$) is the primary of a mid-M-type companion at 14.6\arcsec\ (224 au) that is unlikely to be the source of any acceleration, and HIP 71135 ($\chi^2$=9.3) which has no known stellar companion.  The RV-detected planets are inferred to have (sub)-Neptune-like masses \citep{Feng2019,Stock2020}, and could not produce detectable acceleration.  Finally, we searched the catalog of the Robo-AO M-dwarf Multiplicity Survey \citep{Lamman2020}, which identified candidate companions with separations of 0.1-4\arcsec\ of nearby M dwarfs from the \citet{Lepine2005} catalog.  We identified 20 overlapping stars, none of which had AO-identified companions.

\begin{figure}
    \centering
    \includegraphics[width=\columnwidth]{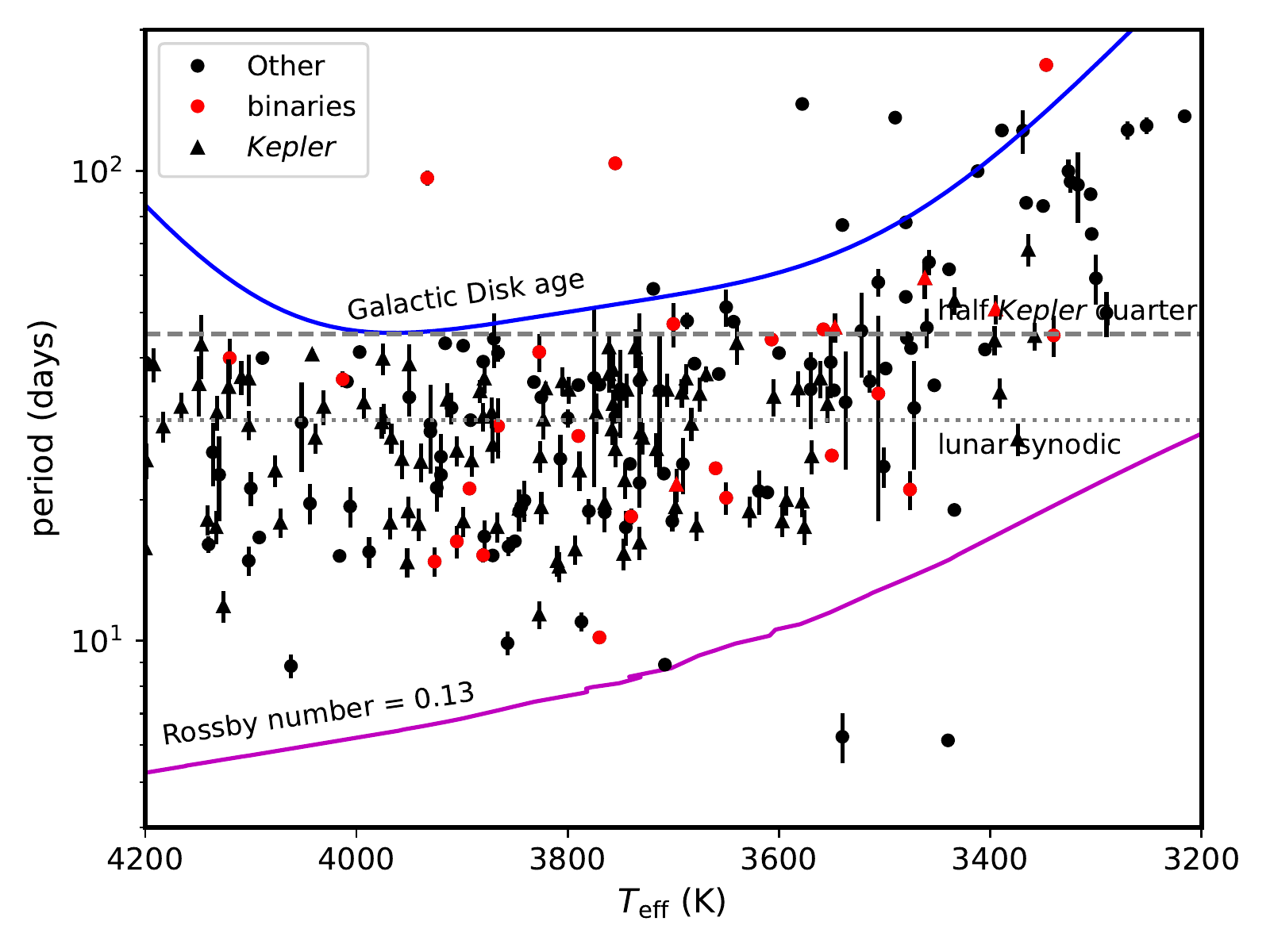}
    \caption{Rotation periods of late-K and early M-type hosts of known exoplanets from the literature and this work.  Members of multi-star systems are indicated in red.  The horizontal dashed and dotted lines mark one half the \kepler\ rotation interval (close to one half the \ktwo\ campaign interval) and the lunar synodic period, near which ground-based detection of \prot\ values are limited.  The magenta curve is the \prot\ above which a star's Rossby number exceeds the critical value (based on convective turnover times from the Dartmouth magnetic model) and activity leaves the ``saturated" phase, and the blue curve is the \prot\ predicted for stars with the age of the Galactic disk using the \citet{Dungee2022} gyrochrone and power-law evolution.  Errors in \teff\ are not shown but are typically 75K.}
    \label{fig:periods}
\end{figure}

Figure \ref{fig:periods} plots \prot\ vs. \teff\  for the catalog of host stars, with \kepler- and non-\kepler\ hosts marked by different points and members of known binaries shown in red.  The horizontal dashed line marks 90 days, the cadence at which \kepler\ performed a role maneuver that introduced systematics in lightcurves, only slightly longer than the 80-day interval which the spacecraft could observe a field during the \ktwo\ mission.  Recovery of rotation periods longer than these intervals is inhibited by these systematics.  The horizontal dotted line is the lunar synodic period of 29.5 days near which the ground-based detection of rotation periods is inhibited.  The magenta curve is the critical \prot\ at which $Ro=0.13$ using the \citet{Jeffries2011} prescription for $\tau_c$.  Nearly all stars are above this line and are thus in the ``unsaturated" regime of dynamo-driven activity, although not necessarily yet in Skumanich-like spin-down.  The blue curve is the expected \prot\ for 10 Gyr, the approximate age of the Galactic disk \citep[][but see \citet{Xiang2022}]{Kilic2017}.    

\section{Age Estimation} \label{sec:ages}

We estimated the age of each star with an established \prot\ (Sec. \ref{sec:catalog} by comparing it to available empirical cluster gyrochrones that (partially) include the \teff\ range of interest, i.e., that of M67 \citet{Dungee2022}, but also that of the 2.7 Gyr-old Ruprecht 147 \citep{Curtis2020}, the 1.4 Gyr-old NGC 752 \citep{Agueros2018}, the 1 Gyr-old NGC 6811 \citep{Curtis2019}, and the 0.67 Gyr-old Praesepe \citep{Douglas2019}, as filtered for binaries and fit with polynomials by \citet{Curtis2020}.   

\citet{Dungee2022} fit the M67 rotation sequence with a fourth-order polynomial with \teff\ over 3200-4200K.
\begin{multline}
\label{eq:gyrochrone}
    P_{\mathrm{4 Gyr}} = 9.66\times10^{-10} \cdot (\teff-4000)^4 + 8.25\times10^{-7} \cdot (\teff-4000)^3\\ + 2.69\times10^{-4} \cdot (\teff-4000)^2  + 0.016 \cdot (\teff-4000) + 25.9,
\end{multline}
\teff\ of M67 members were estimated from their Pan-STARRS $r-i$ colors using a polynomial relation based on synthetic photometry of nearby M dwarf standards \citep{Mann2015}.  \citet{Curtis2020} fit the other rotational sequences as polynomials with \gaia\ $B_P-R_P$ color (see their Appendix B), which we converted to \teff\  using the empirical MS of \citet{Pecaut2013}.  

\begin{figure}
    \centering
    \includegraphics[width=\columnwidth]{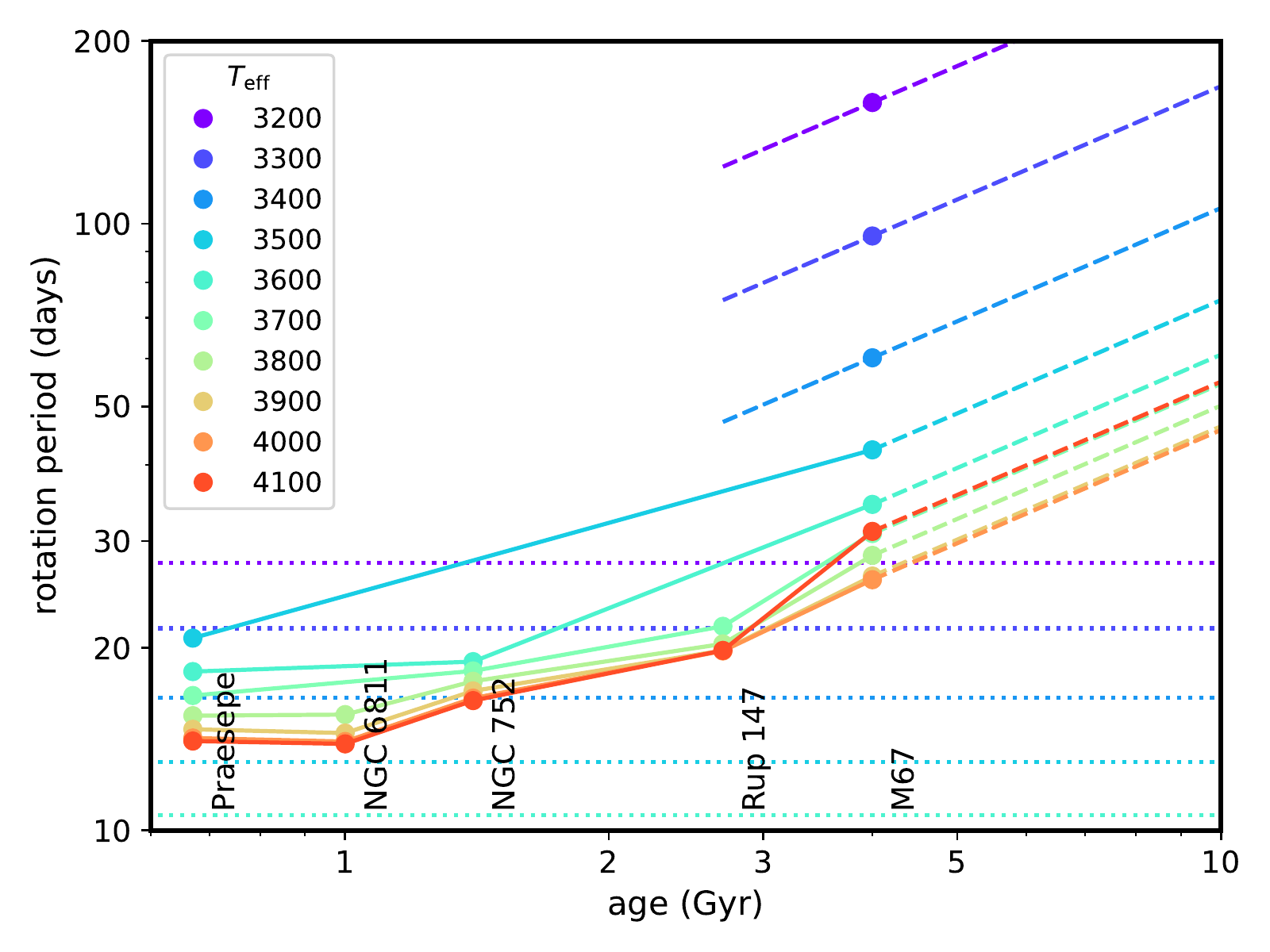}
    \caption{Rotation period vs. age tracks constructed from five cluster gyrochrones (vertically labelled).  The dashed parts of the trajectory are based on the Skumanich-like relation derived by \citet{Dungee2022}, with extrapolation beyond 4 Gyr.  Dotted lines are the periods above which the Rossby number exceeds the critical value for unsaturated activity ($\approx$0.13) at each \teff.}
    \label{fig:gyrochrones}
\end{figure}

We estimated ages by simple power-law interpolation between gyrochrone calibration points (linear interpolation in a log-log plot of period vs. age).  Calibration points were calculated from Eqn. \ref{eq:gyrochrone} and \citet{Curtis2020} for a given stellar \teff.   Figure \ref{fig:gyrochrones} plots the derived period vs. age tracks for a range of representative \teff.   The flattening of the tracks at ages younger than Ruprecht 147 could be ``stalling" of spin-down as a faster rotating radiative core adds AM to the convective envelope \citep{Curtis2020} (see Sec. \ref{sec:summary}).  The bunching of the hotter tracks reflects the weak dependence of \prot\ on \teff\ among co-eval K dwarfs \citep{Curtis2020}.    

A calibration point was used only if (1) \teff\ falls within the range of a gyrochrone; (2) for the particular \teff\ and gyrochrone rotation period, the star would not be in the ``saturated" phase of activity (a condition for a rotational sequence); and (3) the star would not be on the PMS and still affected by contraction and spin-up at that gyrochrone age.  The condition for saturated activity is a Rossby number $Ro =$\prot/\tauconv where $\tau_c$ is the local convective turnover time, to be less than a critical value.  We adopted the relation of $\tau_c$  with luminosity of \citet{Jeffries2011} and critical $Ro$ of 0.13 \citep{Wright2018}.  PMS durations were taken from the Dartmouth standard models of stellar evolution \citep{Dotter2008}.

If the only available calibration point was that of M67 (4 Gyr) then we calculate the age of the star using the Skumanich-like spin-down law that \citet{Dungee2022} found by comparing the rotational sequence of M67 with that of 2.7 Gyr-old Ruprecht 147 \citep{Curtis2020}
\begin{equation}
\label{eq:age}
    t = t_0 \left(P/P_0\right)^{1/n},
\end{equation}
with $n = 0.62$.  However, if the age derived in this manner was $<$2.7 Gyr (i.e., that of Ruprecht 147) we took this to be an upper limit.  If additional calibration points were available we derived an age as above; if that was $<$4 Gyr we then consequently derived ages by power-law interpolation between successfully younger pairs of neighboring calibration points until the derived age lay between the ages of the gyrochrones.  However, if the age determined from the 4 Gyr and next youngest calibration points was $>$4 Gyr, we adopted that age.  This allowed for the occasional pathological cases where the \citet{Dungee2022} gyrochrone produced an age that was slightly $<$4 Gyr, but the gyrochrone pair produced an age slightly $>$4 Gyr.  If no self-consistent age was produced (i.e., the age was younger than the youngest calibration point) we adopted the age of the youngest valid gyrochrone as an upper limit.   

Monte Carlo (MC) realizations of these calculations were performed to estimate the uncertainty in the age, incorporating error in rotation period, stellar parameters, the gyrochronology, and initial conditions (see Sec. \ref{sec:error}).  The mean and standard error of the age distributions were adopted as the nominal age and its uncertainty reported in Table \ref{tab:time}.  If the MC realizations produced only upper limits, the 95 percentile value of the distribution was reported as an upper limit.  If neither values nor upper limits were produced, or the resulting age would place the star on the PMS, no age was assigned.

\section{Age Error Analysis}  \label{sec:error}

Error in gyrochronologic age assignment arises from (1) the formal uncertainty in \prot, as well as systematic errors in  \prot\ not included in the formal error (i.e., aliasing and confusion with harmonics); (2) the formal error in the gyrochrones as well as uncertainty in the ages of the calibrator clusters; (3) errors in \teff\ used to apply the calibration; (4) variation in initial conditions, i.e the angular momentum or rotation rate at an early time; and (5) stellar rotational evolution that deviates from the assumed behavior due to differences in the internal structure of a star or the wind torque.  

\citet{Epstein2014} quantified the effect of uncertainty in \prot\ (\#1 above) and scatter in rotation period at a fixed age (\#4) by projecting ``initial" rotation conditions found in the $\sim$500 Myr-old cluster M37 forward in time with a rotational model.  They found that the age uncertainty among middle-aged M dwarf stars is very large, dominated by the scatter in initial conditions, and as a result the age precision is very poor (factors of two or more between minimum and maximum ages) because, essentially, these stars have not yet spun down to form a rotational sequence, but this depends on the details of the model, including the assumed braking law.     

Here we use Monte Carlo methods to calculate probabilistic distributions of ages --- an approach used by \citet{Otani2022} --- and take the standard deviation as the error.  Where data are available we determine variation empirically; otherwise we rely on models of stellar interior and rotational evolution.     

\subsection{Period Error}
\label{sec:period_error}

We report and use the formal uncertainties in rotation period either from the literature, or from our periodogram analysis.  In the latter case, Gaussian functions are set to the periodogram peak and the standard deviation is taken to be the Gaussian width $\sigma$.  This uncertainty can arise from the finite observation baseline relative to the rotation period, particularly for older stars with longer rotation periods. Significant evolution in the spots causing rotational variability can occur over a few rotation periods \citep{Basri2022}, causing drift in the phase of the overall photometric signal and broadening of the peak.  Differential rotation, combined with changes in the latitude of spots, will also broaden a periodogram peak or even produce distinct neighboring peaks \citep[e.g.,][]{Reinhold2013,Balona2016}.

The first (upper) harmonic might be confused for the true period as a result of the distribution of star-spots \citep[e.g.,][]{Suto2022}, causing the star to appear much younger than it is.  Ground-based observations may incur much larger systematic error as a result of aliasing caused by sampling bias on nocturnal, lunar synodic, or seasonal timescales.  This can distribute the power in single periodic signal into multiple weaker peaks in a periodogram \citep{VanderPlas2018}.  These peaks can either be ``failure modes" if the peak is far from the true period, or if close to the true period and unresolved due to limiting sampling, broaden the peak and resulting uncertainty.  Given the diverse data sets and sources, we point out but do not attempt to quantify such errors in this work.

\subsection{Gyrochrone error}
\label{sec:gyro_error}

\begin{figure}
    \centering
    \includegraphics[width=\columnwidth]{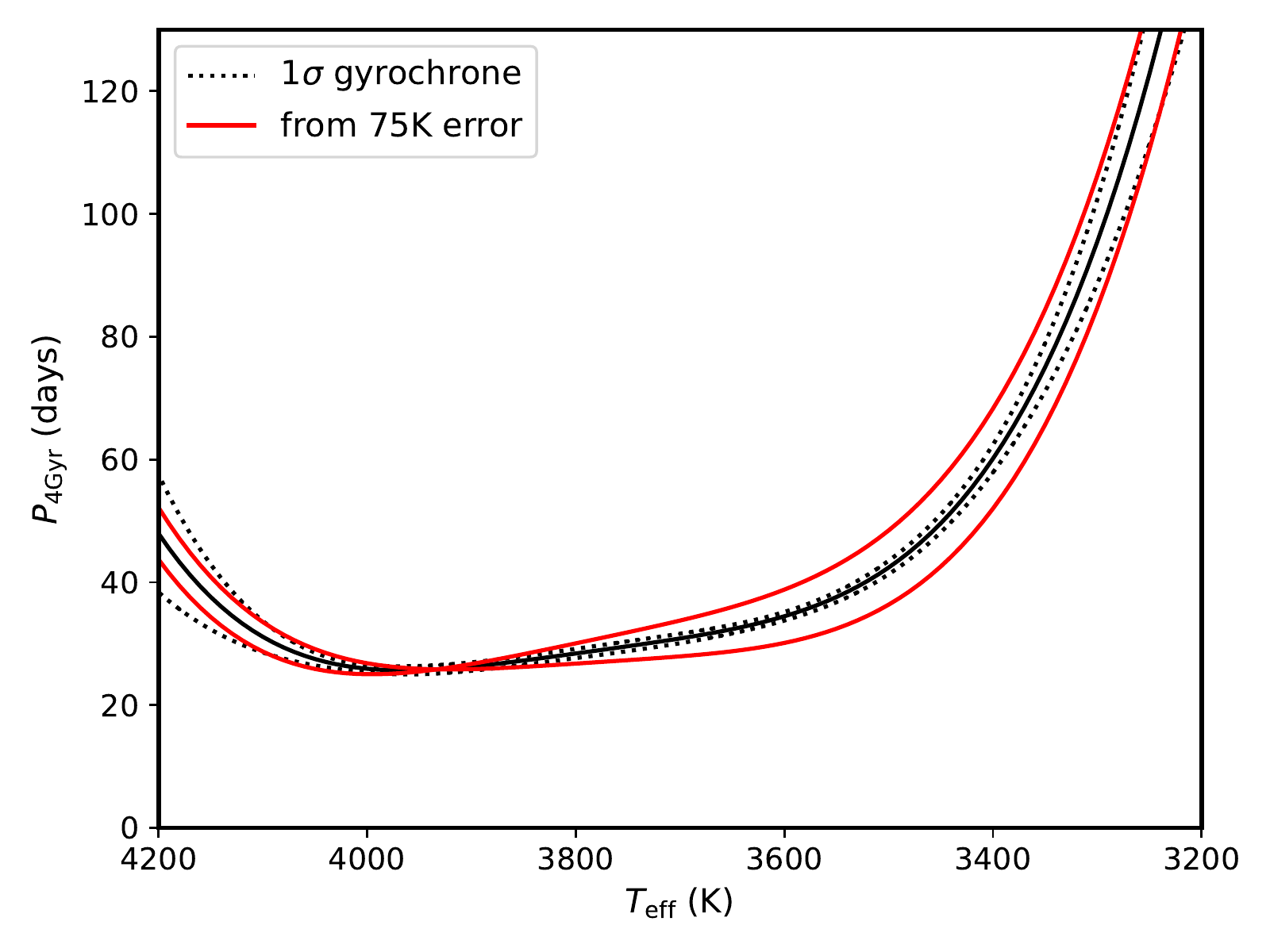}
    \caption{Intrinsic dispersion in the 4 Gyr-old M67 gyrochrone (dashed black lines) established by \citet{Dungee2022} and dispersion produced by errors in \teff\ of 75K (red lines).}
    \label{fig:gyro_mc}
\end{figure}

Due to finite sample size and error in \teff\ and \prot, the gyrochrones themselves have uncertainties.  These are not typically published, but since most of the ages in TIME-Table rely heavily on the 4 Gyr-old M67 gyrochrone, we determined uncertainties in the best-fit polynomial coefficients using Monte-Carlo simulations.  This is shown as the dashed black lines in Fig. \ref{fig:gyro_mc}.  The ages of the clusters used to calibrate the gyrochronology themselves have significant standard errors and systematics; full consideration of these is beyond the scope of this work but should be included when considering rotation-based ages in an absolute sense \citep{Sandquist2021}. 

\subsection{Error in effective temperature}
\label{sec:teff}

Effective temperature \teff\ is usually used as the independent variable of a gyrochronology since it is set by radiated energy per unit area and is thus related to the eddy velocity and magnetodynamo strength in the convective envelope of a star.  Precise and accurate determination of \teff\ is a classic problem in stellar astrophysics, and a particularly acute issue for the coolest stars.  Recent calibrations of the temperature scale of M dwarfs based on interferometric measurement of stellar radii, (\gaia) parallaxes, and the Stefan-Boltzmann law \citep{Boyajian2012,Mann2015} permit precision as good as 50\,K, but most host stars have \teff\ with precision no better than 75\,K.  The M67 gyrochrone of \citet{Dungee2022} was fit to the $g$-$i$ colors of stars, and converted to \teff\ using the temperature scale of \citet{Mann2015}.  This combined with the slope of the M67 gyrochrone will produce formal uncertainties of 0-17\% in age, with the largest value at the cool (3200K) end, decreasing to zero near 3900K, and rising to 12\% at 4200K.   

\subsection{Initial Rotation}
\label{sec:init_rotation}

A major contributor to age error is departure of a star's behavior from a rotation-age relation due to variation in initial conditions, i.e angular momentum/rotation rate at an early age. A star that is initially spinning slower/faster than the mean of the population used to construct the  gyrochronology will have an estimated age that is erroneously older/younger.  Surveys of star-forming regions and very young clusters show that low-mass members emerge from their disk-hosting phase with a wide range of rotation rates at a given mass \citep{Cody2010,Rebull2016,Venuti2017,Rebull2018,Kounkel2019,Serna2021}, perhaps due to variations of disk lifetime brought on by differences in the tidal and ultraviolet environment of stars \citep{Kraus2012a,Roquette2021}.  Since it is not possible to know the initial rotation of an individual star, this variation can induce significant uncertainty in derived ages.  \citet{Epstein2014} used rotational evolution models to show that this greatly limited the utility of rotation-based ages among cooler, older dwarfs. 

The scatter in initial rotation rates can be estimated from observations of the PMS members of very young clusters.  If AM loss through winds is small over the PMS interval, the fractional dispersion in rotation rate at a given mass should be conserved even as the stars contract and spin up.  The fractional dispersion is also conserved during the saturated phase of magnetic activity because the torque is proportional to rotation rate and the spin-down is exponential with a rate-independent time constant.  In the 130 Myr-old Pleiades, no well-defined rotational sequence exists for our range of \teff/colors but there is a strong trend of decreasing \prot\ (6--0.6 days) with decreasing \teff\ \citep{Rebull2016}.  An iterative fit to this over the equivalent de-reddened $V-K$ color range (3.22--5.29) yields a \emph{fractional} dispersion of 45\% (see also Fig. 18 in \cite{Stauffer2016}), but this value does not reflect the numerous outliers, some of which may be binaries.  Likewise, the rotation periods of members of the $\sim$150 Myr-old cluster NGC 2516 \citep{Fritzewski2020} within this same color range have a scatter of $\gtrsim$50\%, about one order of magnitude larger than what is observed at later epochs.    

This large scatter in initial (ZAMS) rotation rates $P_0$ will be compressed once the stars decelerate to the unsaturated regime at $P_{\rm crit} \sim 0.13 \tau_c$ \citep{Wright2018} where the torque becomes highly rotation rate-dependent and period-time trajectories flatten.  In the saturated regime, $P(t) = P_0 e^{t/T}$, where $T$ is the spin-down timescale \citep{Matt2015}, thus a star with a ZAMS rotation period that differs by $\Delta P_0 \ll P_0$, will reach $P_{\rm crit}$ at a time that differs by 
\begin{equation}
\label{eq:delta_t}
\Delta t = - T \log \left(\frac{P_0 + \Delta P_0}{P_0}\right). 
\end{equation}
Because subsequent rotational evolution proceeds from the condition $P = P_{\rm crit}$, independent of $P_0$, the star will experience the same rotation history, but delayed by $\Delta t$.  Thus this is the error in gyrochronological age induced by $\Delta P_0$.  $T$ depends on the braking torque parameter and moment of inertia of the star.  \citet{Somers2017} found that stars in mass bins of $0.25-0.40$\msun\ and $0.40-0.60$\msun\ lose 25\% and 39\%, respectively, of their AM in the $\approx115$ Myr interval between the age of the Upper Scorpius star-forming region and the Pleiades cluster, implying a spin-down timescale of 225-400 Myr.  \citet{Somers2017} also found a dispersion of 0.21 and 0.44 dex in the specific (mass-normalized) AM of Pleiades stars for stars in these respective mass bins.  Application of Eqn. \ref{eq:delta_t} produces an age error of about 200 Myr, which for a nominal 4 Gyr-old star is 5\%.    

\begin{figure*}
    \centering
    \includegraphics{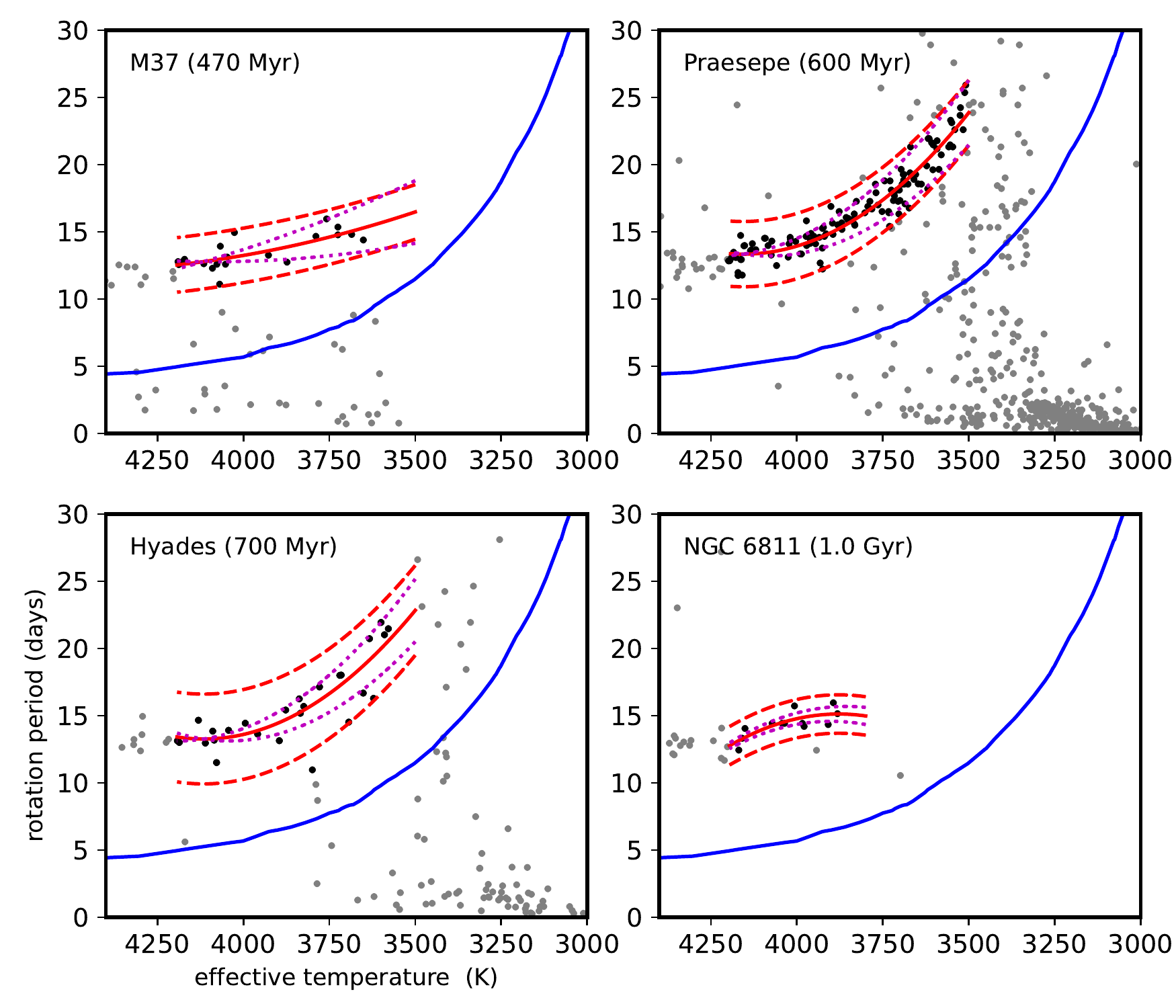}
    \caption{Fits of the cool dwarf rotational sequences of four open clusters.  (The existence of a rotation sequence in M37 M dwarfs is unclear).  Black points indicate stars used in iterative, outlier-rejection fits.  The solid red line is the fit, the dashed red lines are the $\pm2.5\sigma$ rejection boundaries, and the dotted magenta lines show the extent of the scatter induced solely by an error of \teff\ of 75K. The blue line is the critical rotation period for $Ro=0.13$ using the relation between convective turnover time and luminosity of \citet{Jeffries2011}, with luminosities from the Dartmouth stellar evolution models \citep{Dotter2008,Feiden2016}.  Stars below this line will have saturated magnetic fields and experience exponential spin-down.}
    \label{fig:rotation_fits}
\end{figure*}

Finally, the intrinsic dispersion in the rotation sequences of cluster stars can be used to empirically estimate age error due to rotation rate dispersion.   For stars undergoing power-law spin-down \citep[\prot $\propto t^{\chi}$,][]{Skumanich1972}, the dispersion in age consistent with a given \prot\ is related to the dispersion in period for a given age (i.e., in the co-eval population of a cluster):
\begin{equation}
\label{eq:delta_p}
\frac{\Delta t}{t} = \frac{1}{\chi} \frac{\Delta P}{P} 
\end{equation}
We measured the outlier-excluded dispersion $\Delta P$ around the best-fit rotational sequence of M67 to be 1.6 days, but this is consistent with the 10\% measurement errors alone, and the intrinsic dispersion could be smaller.  The dispersion in other, younger ($\lesssim$1 Gyr) clusters like those observed by \ktwo\ can be better determined, but with the caveat that Skumanich-like spin-down only applies to times much later than the era of saturated magnetic activity.  A well-defined rotational sequence is apparent in the Praesepe cluster (upper right-hand panel of Fig. \ref{fig:rotation_fits}), estimated to be 600 Myr-old \citep{Gossage2018}.   We fit a second-order polynomial with iterative 3$\sigma$ outlier rejection to 3500-4200K members with rotation periods cataloged by \citet{Douglas2019}.  (Cooler stars have yet to converge to an identifiable sequence.)  The scatter is 0.97 days (N=121).  The same analysis applied to the much sparser Hyades sample yields the same dispersion: 0.95 days (N=13), a fractional dispersion of about 5\% (bottom left panel of Fig. \ref{fig:rotation_fits}).  The data on M dwarfs in other, older clusters are much sparser: \citet{Curtis2019} estimated a dispersion of $\pm$10\% for the coolest stars in NGC 6819 ($\approx$2.5 Gyr) and Ruprecht 147 ($\approx$2.7 Gyr) but of these, only \emph{five} stars have \gaia\ $B_p-R_p$ colors that correspond to our \teff\ range.  Using the \citet{Godoy-RIvera2021} catalog of members of M37 \citep[$\approx$470 Myr][]{Fragkou2022} and NGC 6811 ($\approx$950 Myr) with rotation periods we estimate dispersions of 0.8 days (n=20) and 0.6 days (n=12) respectively (upper left and bottom right panels of Fig. \ref{fig:rotation_fits}).  The existence of a rotation sequence among M37 M dwarfs is not clear.  Some of the observed dispersion in these clusters is probably the result of errors in \teff: for a standard error of 75\,K, this is $\sim$0.5-1 day, depending on cluster and \teff\ (purple dotted lines in Fig. \ref{fig:rotation_fits}).  Thus the actual period dispersion could be substantially lower than the observed values.  

We adopted a conservative fractional dispersion in $P_0$ of 5\%, with the caveat that current data on the establishment of a rotational sequence at these epochs only extends to a spectral type of M2 (\teff $\sim$ 3550\,K).  For $n = 0.62$, this corresponds to an age dispersion of 8\% (Eqn. \ref{eq:delta_p}), which we incorporate in our MC realizations of age estimates by adopting a Gaussian distribution.    These calculations assume that a rotational sequence has developed by the epoch of interest for the relevant \teff, i.e., that stars have spun down into the   unsaturated regime.  This is not the outcome for M dwarfs in the models of \citet{Epstein2014}, which explains their large predicted age errors, but it is the case at least by 4 Gyr for this \teff\ range (see Sec. \ref{sec:caveats}).         

\subsubsection{Metallicity}
\label{sec:metallicity}

The rotational evolution of cool dwarf stars is expected to be metallicity-dependent through the effects of opacity and mean atomic weight on the interior structure and dynamics of the star, which in turn govern the moment of inertia, and the scale of convection that drives the dynamo responsible for star's magnetic field, activity, and mass loss through a wind.  Moreover, at a fixed \teff, a more metal-rich MS star will have a higher mass.   The [Fe/H] of the nearby stellar clusters used to construct gyrochronologies is close to solar:  M67, Ruprecht 147, and NGC 6811 are within uncertainties of solar \citep{Pasquini2008,Bragaglia2018,Molenda-Zakowicz2014}, while both Praesepe and the Hyades have [Fe/H] = +0.15 \citep{Cummings2017}.  However, stars in the solar neighborhood have metallicities ranging from $-1$ to $+0.5$ \citep{Toyouchi2018}, and the M dwarf stars observed by \kepler\ have a [Fe/H] distribution that is approximately Gaussian with a mean of $-0.09$ dex and standard deviation of $\pm$0.22 dex \citep{Gaidos2016}.

We predict the effect of non-solar metallicity on the duration of the stellar PMS phase and the subsequent spin-down on the MS, which we model as exponential ``saturated" spin-down from an initial rotation period $P_0$ to the critical value $P_{\rm crit} = 0.13 \tau_c$  at which the Rossby number exceeds a critical value 0.13 \citep{Wright2018}, and unsaturated, Skumanich-like power-law spin-down thereafter.  The error in an age based on solar-metallicity gyrochrones induced by a non-solar metallicity is the sum of the variation in these two intervals.  We used the Dartmouth standard (non-magnetic) models to compute these for a range of masses and [Fe/H] in 0.1 dex intervals from $-0.7$ to $+0.5$.  The MS saturated spin-down interval is taken to be:
\begin{equation}
\label{eq:ms_interval}
    T_{\rm sat} =  \frac{I}{n} \log \frac{P_{\rm crit}}{P_0},
\end{equation}
where $I$ is the moment of inertia (nearly constant for MS M dwarfs) and $\Gamma$ is the (constant) torque parameter.  Both $I$ and $P_{\rm crit}$ are metallicity-dependent and were calculated using the Dartmouth models.  We adopted the scaling relationship for the torque parameter from \citet{vanSaders2013}, which is based on \citet{Matt2012}:
\begin{equation}
\label{eq:torque_scaling}
    \Gamma \propto R_*^{3.1} L_*^{0.56} M_*^{-0.22} p_{\rm phot}^{0.44} 
\end{equation}
where $p_{\rm phot}$ is the pressure at the photosphere.  $p_{\rm phot}$ will be proportional to gravity and inversely proportional to the specific opacity $\kappa$.  The mass inferred for a given MS \teff\ will also vary with [Fe/H] because the radius and hence luminosity changes.  Using the $M_*-M_{K_s}$ (mass-luminosity) relations of \citet{Mann2019}, and exploiting the fact that the bolometric correction for the $K_s$-band is only weakly dependent on [Fe/H] \citep{Mann2015} and thus will be approximately fixed at a given \teff, $L_* \approx M_*^{2.7}$ in this range.  With that relation and the Stefan-Boltzmann law, at a given mass and \teff, Eqn. \ref{eq:torque_scaling} becomes:
\begin{equation}
\label{eq:torque_scaling2}
    \Gamma \propto R_*^{4.53} \kappa^{-0.44} 
\end{equation}
In the interiors of cool stars where bound-free opacity dominates, the opacity will scale linearly with the metal abundance $Z \propto 10^{[Fe/H]}$.  We calculated for $\Gamma$ for the solar-metallicity case by finding the age at which Skumanich-like rotational evolution marched backward from the M67 gyrochrone \citep{Dungee2022} gives \prot=$P_{\rm crit}$ and then setting $\Gamma$ so that exponential spin-down from \prot=$P_0$ also reach $P_{\rm crit}$ at this age, i.e:
\begin{equation}
\label{eq:torque}
    \Gamma = \frac{I_0}{\rm 4 Gyr}\left(\frac{P_4}{P_{\rm crit}}\right)^{1/0.62}\log \frac{P_{\rm crit}}{P_0}
\end{equation}

The PMS interval of M dwarfs is not readily defined since these stars gradually approach the MS.  Since we are concerned only with the effect of stellar contraction on spin-up, we define the interval at which the timescale for contraction and spin-up (taken to be the logarithmic change in the momentum of inertia with time) greatly exceeds the timescale for spin-down by saturated magnetic activity.  Here, we adopted ``greatly" to be 10$\times$ but our estimates are not sensitive to the exact figure.  

\begin{figure}
    \centering
    \includegraphics[width=\columnwidth]{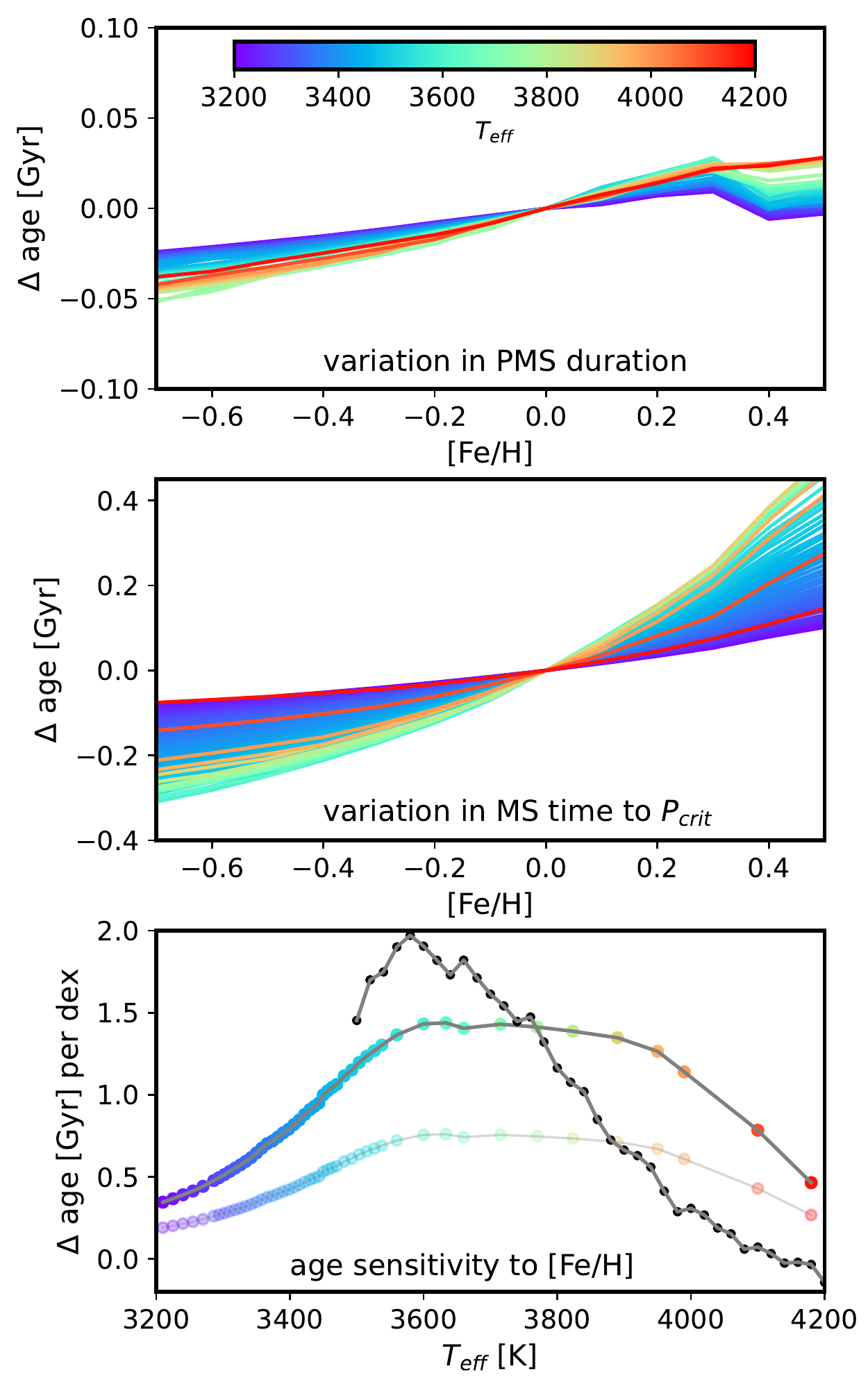}
    \caption{Difference in actual age relative to gyrochrone age due to non-solar metallicity due to variation in the PMS duration (a) and MS interval required for spin-down sufficiently for Ro to exceed the critical value 0.13 and the star to leave the saturated phase of activity and follow power-law Skumanich-like spin-down (middle).   Positive values means that a star will be older than its gyrochronologic age.  Each color corresponds to a different mass track in Dartmouth standard model calculations, converted to \teff\ on the main-sequence using the empirical relation of \citet{Pecaut2013}.  The light colored line in the bottom panel is the slope of the summed intervals vs. [Fe/H] obtained at each value of \teff.   The black curve is the slope calculated from a full model of metallicity-dependent spin-down over a representative age of 4 Gyr.  The heavy colored line is the PMS interval plus twice the MS interval used as an approximation for the actual sensitivity that reproduces the shape and magnitude of the model simulations.}
    \label{fig:metallicity}
\end{figure}

The top panel of Fig. \ref{fig:metallicity} plots the variation in PMS duration relative to the solar-metallicity cases vs. [Fe/H] for the same mass/\teff\ cases as the top panel.  The middle panel of Fig. \ref{fig:metallicity} plots variation of MS $T_{\rm sat}$ relative to the solar-metallicity value vs. [Fe/H] for 40 different mass tracks with MS \teff\ falling within 3200-4200K.  As one metric of the age error due to non-solar [Fe/H] we added the PMS and saturated spin-down durations and performed linear regression with [Fe/H] over the range of $-0.3$ to $+0.3$, where most M dwarfs fall.  This slope of this regression vs. \teff\ is the light-colored curve in the bottom panel of Fig. \ref{fig:metallicity}.  The sensitivity of the age estimates to [Fe/H] increases from 0.05 Gyr/dex at 3200K, peaking at 0.2 Gyr/dex at around 3900K, and declining in the K dwarf regime (Fig. \ref{fig:metallicity}).  This behavior is almost entirely a consequence of the metallicity dependence of the braking torque \citep{vanSaders2013}, with a lesser contribution from changes in mass with metallicity at a fixed \teff.

Although the power-law index of the Skumanich-like spin down observed on long time-scales is not considered metallicity-dependent, metallicity-dependent torque could still impart additional deviation from predictions based on solar-metallicity gyrochrones during the transition to purely power-law behavior.   To quantify this, we performed calculations of rotation evolution using the models of \citet{Claytor2020}, which use the torque scaling of Eqn. \ref{eq:torque_scaling}.  We used the stellar model interpolation and Markov Chain Monte Carlo (MCMC) tools in \texttt{kiauhoku} \citep{kiauhoku} to make \prot-based age estimates of 4 Gyr-old model stars with a given \teff\ and varying [Fe/H], but assuming solar [Fe/H].  We performed a linear regression of the inferred age minus the ``true" age (4 Gyr) versus [Fe/H] for different values of \teff, and the slope is plotted as the black curve in the bottom panel of Fig. \ref{fig:metallicity}).  (These calculations do not go below 3500 K because of incompleteness in the model grid.)  This curve has the same overall shape as our curve of PMS+MS age sensitivity (light colored curve) but is generally larger in magnitude, as expected.

Based on a comparison of the curves in Fig. \ref{fig:metallicity}a, we approximately incorporate the metallicity dependence of spin-down during the transition to pure Skumanich-like behavior by doubling the offset during the MS saturated and adding it to the PMS offset.  This is the heavy colored curve in the bottom panel.  We multiply the slope by a typical uncertainty of $\pm$0.1 dex in [Fe/H] and add this (up to $\pm$150 Myr) to our error budget.   We also use \texttt{kiauhoku} to calculate individual [Fe/H]-dependent \emph{corrections} for the age of each star with known metallicity and \teff $>$3500K that can be added to age estimated from our solar-metallicity gyrochronology.    

\section{Results: Planet Host Star Ages}

\begin{figure}
    \centering
    \includegraphics[width=\columnwidth]{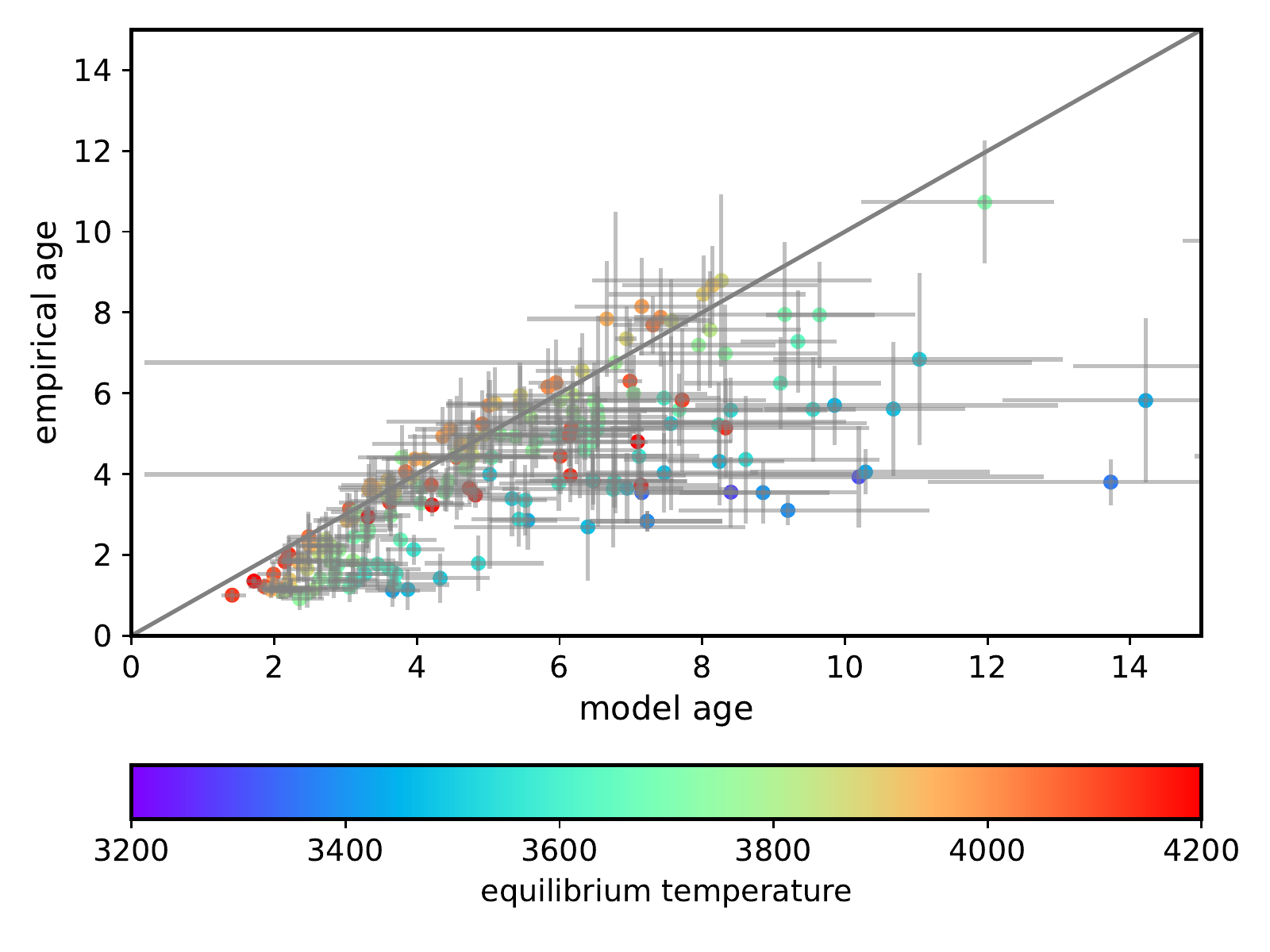}
    \caption{Empirical ages of host stars based on the rotation-age relations in Fig. \ref{fig:gyrochrones} vs. the estimates using the \texttt{kiauhoku} model \citep{kiauhoku}.}
    \label{fig:model-empirical}
\end{figure}

Table \ref{tab:time} provides the \teff\ and [Fe/H] (if available) that were used for the gyrochrone calculations, the rotation period, the method and instruments used to obtain it and the reference, and the estimated age and uncertainty.  If a metallicity is available, we also provide, but do not incorporate, a \texttt{kiauhoku}-calculated value for the [Fe/H]-dependent correction.  Only the first 50 entries are shown; the complete machine-readable table is provided on Zenodo (DOI: 10.5281/zenodo.7578269).  Figure \ref{fig:model-empirical} compares the empirical ages of host stars to ages using generated with the \texttt{kiauhoku} model \citep{kiauhoku}.  There is good agreement among the warmer stars in the sample, but a clear trend of older \texttt{kiauhoku}-based estimates for cooler \teff, where the models have not been calibrated. The systematic offset of model-derived ages for the coolest stars further illustrates the need for calibrators across the full range of temperature and age.

\begin{figure}
    \centering
    \includegraphics[width=\columnwidth]{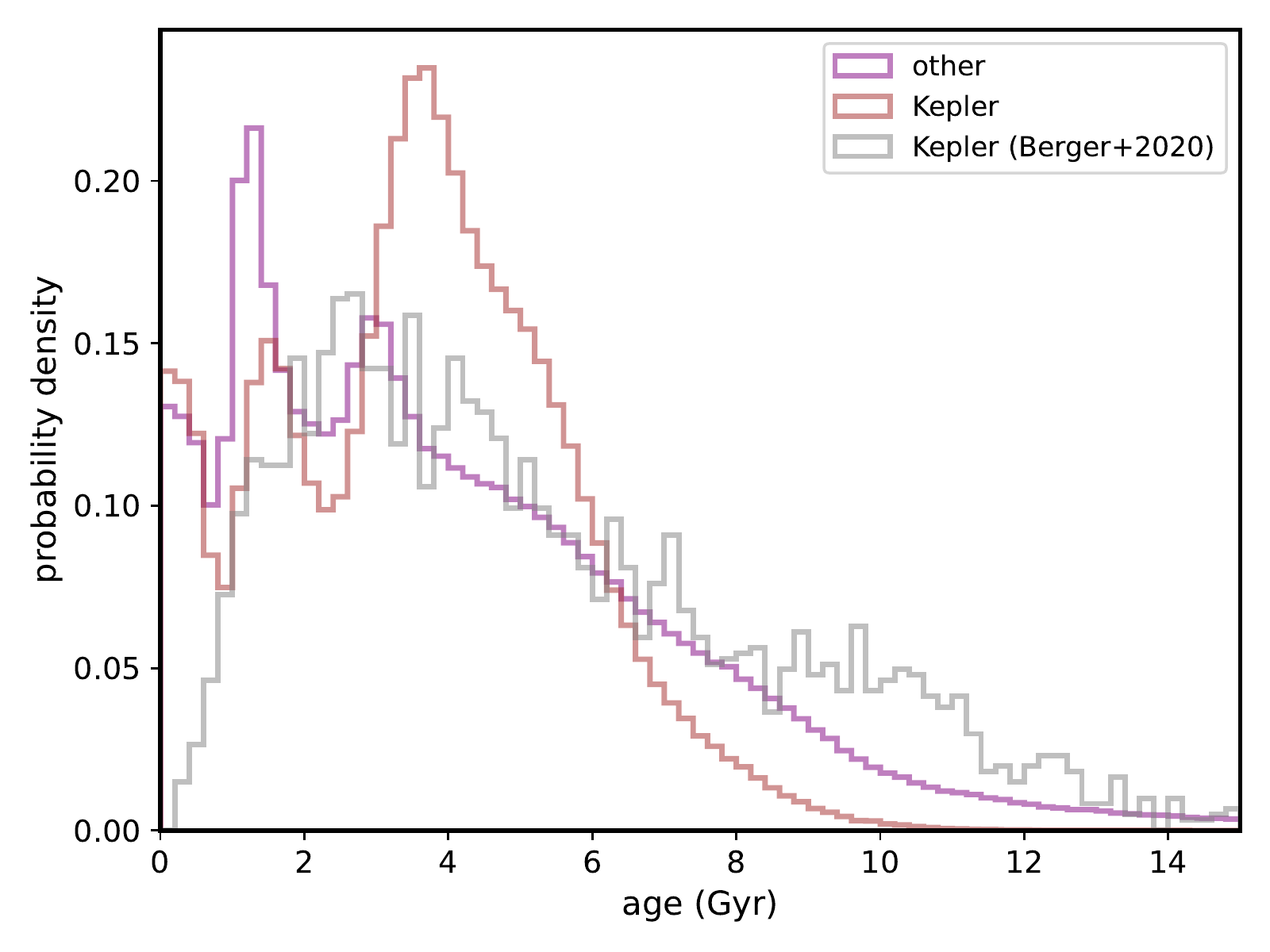}
    \caption{Distribution of estimated ages for KOIs and all other known M dwarf host stars, accounting for the uncertainties.  The grey curve is the isochrone-based age distribution for all \kepler\ stars from \citet{Berger2020}.}
    \label{fig:age_dist}
\end{figure}

The distribution of ages assigned to Monte Carlo realizations is plotted in Fig. \ref{fig:age_dist}, where we have plotted the KOIs and all other host stars with separate curves as distinct in terms of sensitivity and systematics as well as (potentially) stellar populations, and compare these to the isochrone-based distribution for all \kepler\ stars from \citet{Berger2020}.  For upper limits, ages were drawn from a uniform distribution from zero to the upper limit.  We did not exclude known binary stars from these distributions since the effect of binaries depends on semi-major axis in a manner that is still being actively investigated \citep[e.g.,][]{Messina2019}.   

\begin{figure}
    \centering
    \includegraphics[width=\columnwidth]{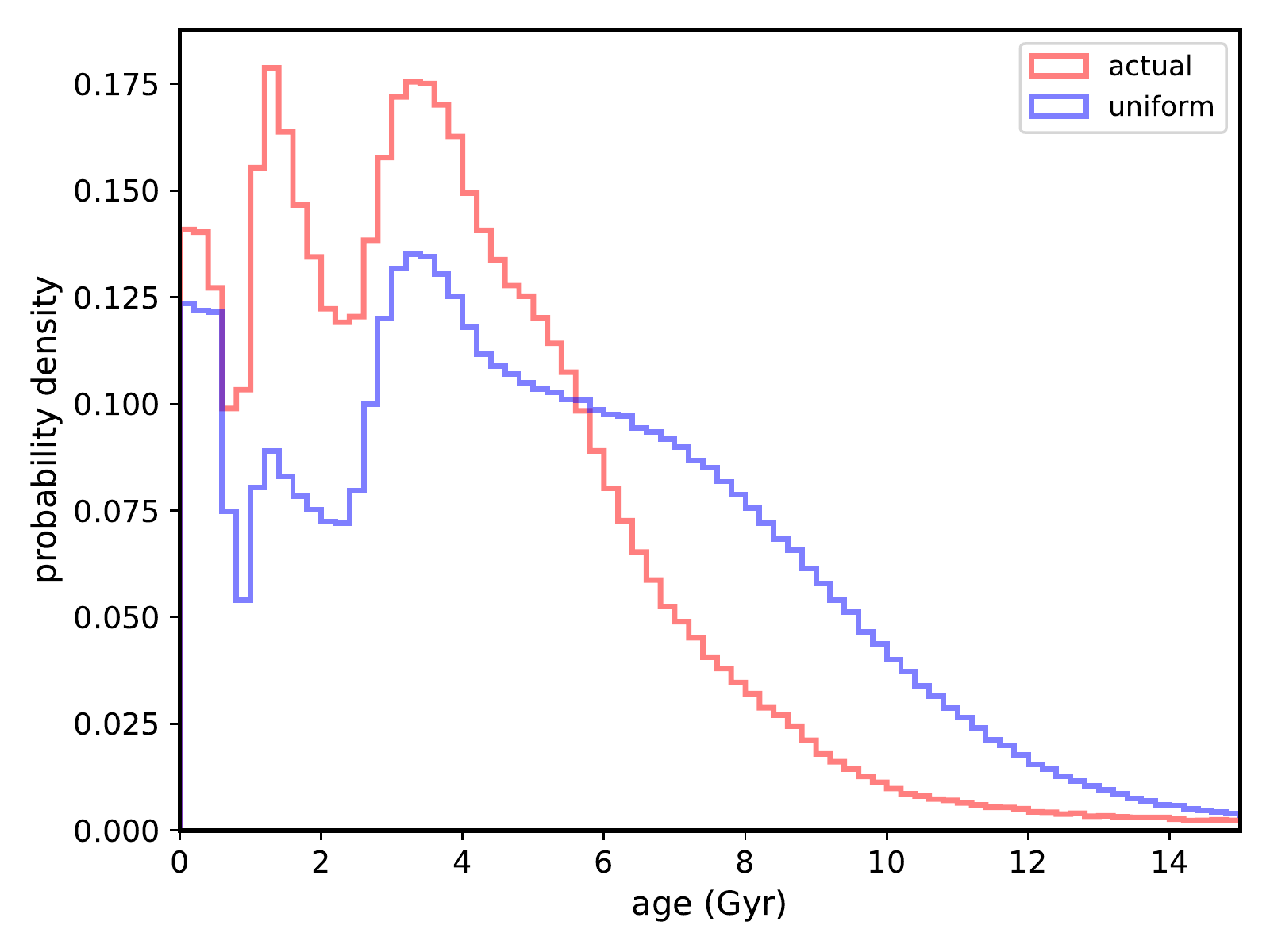}
    \caption{Distribution of actual ages vs. those inferred from a ``mock" population of stars with the same \teff\ distribution as the exoplanet host stars, but a uniform 0-10 Gyr age distribution (see text for details).}
    \label{fig:mock}
\end{figure}

To help discern between actual structure and systematics in the age distribution, we created a mock stellar population with a uniform age distribution, \teff\ drawn with replacement from the actual catalog, and \prot\ calculated with a simple model of the spin-down of stellar population.  The model assumed solid-body rotation and power-law spin-down with index $\gamma = 0.62$ for $Ro > 0.13$, i.e., at MS ages $t > t_{\rm crit}$ where the condition \prot\ $> P_{\rm crit}$, as defined before, is satisfied.  For main-sequence ages $t < t_{\rm crit}$, stars undergo exponential spin-down with a time constant set such that at $t = 0$, \prot\ is equal to an initial value $P_0$.  The initial period is derived from the specific momentum distribution of among $\sim$10 Myr-old M dwarf members of the Upper Scorpius star-forming region \citep{Somers2017}, and the moments of rotational inertia from the Dartmouth stellar evolution models.  Main sequence ages were a random draw from a uniform 0-10 Gyr distribution, minus the PMS duration as taken from Dartmouth solar-metallicity models \citep{Dotter2008}.  (The \prot\ of PMS stars is fixed at $P_0$, but this choice is not important since these stars were subsequently excluded.)  We added the Gaussian-distributed error of 7.4\% to the periods, the median of the distribution of actual error, and then derived the ages and errors with the same routines used for the actual exoplanet host star catalog.  The actual and mock distributions are compared in Fig. \ref{fig:mock}.

There is a marked deficit and marked structure at $<$3 Gyr in the age distribution of stars in both the \kepler\ and non-\kepler\ samples (Fig. \ref{fig:age_dist}), but this feature also appears in the simulation of a uniform distribution of age (Fig. \ref{fig:mock}), showing that the inferred age distribution is heavily affected by systematics, i.e. the incomplete working gyrochronology over the entire \teff\ range at young ages.  Monte Carlo realizations that fall in these gaps are assigned upper limits and thus not correctly represented in this distribution.  Both the inferred \kepler\ and non-\kepler\ distributions decline with age, and more rapidly than that inferred from the mock uniform-age population.  A declining rotation-based age distribution was also inferred for solar-type \kepler\ host stars and is in part a bias caused by the  difficulty of detecting the lower amplitude, longer period rotational signals that are more prevalent around older stars \citep{Walkowicz2013}.  This pattern is also mimicked by an age distribution of \kepler\ target stars based on isochrone analysis \citep[][grey curve in Fig. \ref{fig:age_dist}]{Berger2020}; this is also biased towards younger (and more massive stars) that evolve more quickly.  The \kepler\ distribution peaks at older ages than the non-\kepler\ sample, which could be due to the greater sensitivity and longer monitoring interval of the prime mission, but perhaps also because \kepler\ was observing a field centered at $b = 13$ deg containing a slightly older population further above the Galactic plane.  The \kepler\ distribution terminates at 10~Gyr, about the age of the Galactic disk.  The distribution of non-\kepler\ host stars has a tail that extends well beyond 10~Gyr but this is largely due to host stars with significant uncertainties in \prot, along with a handful of binaries (see below).

\begin{table*}
\label{tab:time}
\caption{TIME-Table: Catalog of Cool Host Stars with Established Rotation Periods.$^{1}$}
\begin{tabular}{lllllllll}
\hline
Name & \teff & Fe/H & Period (unc) & Method$^{2}$ (Instruments$^{3}$) & Reference & Age (unc) & Corr$^{4}$ & Note$^{5}$\\
 & [K] & [dex] & [days] &  &  & Gyr & Gyr & \\
\hline
EPIC 201170410 & 3650 & -0.05 & 20.16 (1.6)  & P (K2) & this work & 1.77 (0.58) & +0.00 & binary\\
EPIC 211822797 & 3856 & +0.14 & 15.88 (0.72)  & P (K2) & \cite{Reinhold2020} & 1.18 (0.24) & +0.03 & \\
G 264-012 & 3326 & +0.10 & 100.0 (6.0)  & PS (TE/ME/AN/CA) & \cite{Amado2021} & 6.66 (3.26) & +0.12 & \\
G 9-40 & 3713 & +0.04 & 34.08 (11.46)  & P (K2) & \cite{Reinhold2020} & 5.35 (2.57) & +0.00 & \\
GJ 1132 & 3270 & -0.12 & 122.3 (5.5)  & P (ME) & \cite{Cloutier2017} & 6.31 (3.23) & ----- & \\
GJ 1148 & 3304 & +0.16 & 73.5 & P (HA) & \cite{Hartman2011} & 4.44 (1.6) & +0.51 & no error\\
GJ 1214 & 3252 & +0.29 & 125.0 (5.0)  & P (ST) & \cite{Mallonn2018} & 5.91 (3.0) & ----- & \\
GJ 1252 & 3458 & +0.10 & 64.0 (4.0)  & P (WA) & \cite{Shporer2020} & 6.61 (2.34) & +0.58 & \\
GJ 1265 & 4052 & -0.04 & 29.17 (6.27)  & P (K2) & \cite{Reinhold2020} & 4.44 (1.53) & +0.11 & \\
GJ 15 A & 3607 & -0.32 & 43.82 (0.56)  & PS (Fa/HI) & \cite{Howard2014} & 5.87 (1.14) & -0.66 & binary\\
GJ 176 & 3680 & +0.14 & 38.92 & P (AS) & \cite{Kiraga2007} & 5.6 (0.9) & +0.52 & no error\\
GJ 229 & 3790 & ----- & 27.3 & P (AS) & \cite{Suarez-Mascareno2016} & 3.77 (0.51) & ----- & binary,no error\\
GJ 251 & 3389 & -0.03 & 122.1 (2.05)  & PS (WA/CA) & \cite{Stock2020} & 12.52 (5.86) & ----- & \\
GJ 27.1 & 3578 & ----- & 139.0 (3.5)  & P (AS) & this work & 34.94 (7.93) & ----- & too old\\
GJ 273 & 3317 & +0.09 & 93.5 (16.0)  & S (HA) & \cite{Suarez-Mascareno2017} & 6.12 (3.17) & +0.09 & \\
GJ 3138 & 3899 & -0.30 & 42.5 & S (HA) & \cite{Astudillo-Defru2017b} & 8.44 (0.95) & +0.14 & no error\\
GJ 3293 & 3600 & +0.11 & 41.0 & S (HA) & \cite{Astudillo-Defru2017b} & 5.19 (1.06) & +0.40 & no error\\
GJ 338 B & 3770 & -0.03 & 10.17 & P (K2) & \cite{Magaudda2020} & <0.91 & ----- & binary,no error\\
GJ 3470 & 3611 & +0.20 & 20.7 (0.15)  & P (Fa) & \cite{Biddle2014} & 1.65 (0.54) & +0.16 & \\
GJ 3473 & 3347 & +0.11 & 168.3 (3.65)  & P (ME/TJ) & \cite{Kemmer2020} & 16.46 (8.46) & ----- & binary\\
GJ 357 & 3480 & -0.12 & 77.8 (2.1)  & P (AS/AN/NS) & \cite{Luque2019} & 9.76 (3.31) & -0.30 & \\
GJ 367 & 3687 & -0.01 & 48.0 (2.0)  & P (WA) & \cite{Lam2021} & 7.95 (1.31) & +0.04 & \\
GJ 3779 & 3324 & +0.00 & 95.0 (5.0)  & PS (ME/CA) & \cite{Luque2018} & 6.13 (2.87) & ----- & \\
GJ 3929 & 3369 & +0.00 & 122.0 (13.0)  & P (HA/AN) & \cite{Kemmer2022} & 11.1 (5.76) & -0.36 & \\
GJ 393 & 3548 & -0.18 & 34.15 (0.22)  & P (K2) & \cite{Amado2021} & 3.32 (0.87) & -0.24 & \\
GJ 3942 & 3850 & -0.04 & 16.3 (0.1)  & PS (AS/HA) & \cite{Suarez-Mascareno2018} & 1.18 (0.24) & +0.01 & \\
GJ 3998 & 3825 & -0.16 & 33.0 & P (AP) & \cite{Giacobbe2020} & 5.25 (0.69) & ----- & no error\\
GJ 411 & 3719 & -0.36 & 56.16 (0.27)  & P (Fa) & \cite{Diaz2019} & 10.73 (1.52) & -0.52 & \\
GJ 414 A & 4120 & +0.24 & 40.0 (4.0)  & P (KE) & \cite{Dedrick2021} & 5.87 (1.77) & +0.51 & binary\\
GJ 4276 & 3440 & +0.12 & 6.14 & P (AP) & \cite{Giacobbe2020} & --- & ----- & no age,no error\\
GJ 436 & 3479 & +0.01 & 44.1 (0.2)  & P (Fa) & \cite{Lothringer2018} & 4.32 (1.09) & +0.00 & \\
GJ 486 & 3290 & -0.15 & 49.9 (5.5)  & P (AS/LC/WA/TJ/OS) & \cite{Caballero2022} & 3.51 (0.81) & -0.13 & \\
GJ 514 & 3755 & -0.09 & 30.0 (0.9)  & S (HA) & \cite{Suarez-Mascareno2017} & 4.11 (0.6) & +0.03 & \\
GJ 581 & 3490 & -0.09 & 130.0 (2.0)  & S (HA) & \cite{Robertson2014} & 23.26 (7.7) & ----- & \\
GJ 625 & 3540 & -0.35 & 76.79 (0.13)  & P (WA) & \cite{Diez-Alonso2019} & 11.96 (3.14) & ----- & \\
GJ 628 & 3305 & -0.02 & 89.3 (1.8)  & P (Fa) & \cite{Kane2017} & 5.27 (2.32) & ----- & \\
GJ 649 & 3741 & +0.03 & 23.8 (0.1)  & P (AS) & \cite{Diez-Alonso2019} & 2.96 (0.51) & +0.11 & \\
GJ 667 C & 3755 & ----- & 103.9 (0.7)  & S (HA) & \cite{Suarez-Mascareno2015} & 30.59 (4.06) & ----- & binary,too old\\
GJ 674 & 3453 & -0.28 & 35.0 (0.1)  & P (AS) & \cite{Suarez-Mascareno2016} & 2.85 (0.72) & -0.28 & \\
GJ 676 A & 3827 & +0.23 & 41.2 (3.8)  & S (HA) & \cite{Suarez-Mascareno2015} & 7.58 (1.46) & +0.56 & binary\\
GJ 685 & 3844 & +0.10 & 19.3 (0.3)  & P (TJ) & \cite{Diez-Alonso2019} & 2.25 (0.36) & +0.15 & \\
GJ 687 & 3439 & +0.05 & 61.8 (1.0)  & P (Fa) & \cite{Burt2014} & 5.82 (1.99) & +0.42 & \\
GJ 720 A & 4013 & +0.01 & 36.05 (1.41)  & S (HA) & \cite{Gonzalez-Alvarez2021} & 6.26 (1.05) & +0.21 & binary\\
GJ 740 & 3832 & +0.08 & 35.56 (0.07)  & S (CA/HA) & \cite{Toledo-Padron2021} & 5.97 (0.72) & +0.32 & \\
GJ 832 & 3522 & -0.30 & 45.7 (9.3)  & S (HA) & \cite{Suarez-Mascareno2015} & 5.2 (2.13) & -0.98 & \\
GJ 849 & 3551 & +0.25 & 39.2 (6.3)  & S (HA) & \cite{Suarez-Mascareno2015} & 4.33 (1.55) & +0.68 & \\
GJ 876 & 3472 & +0.17 & 31.31 (8.15)  & P (K2) & \cite{Reinhold2020} & 2.7 (1.35) & +0.29 & \\
GJ 96 & 3892 & +0.14 & 29.5 (0.5)  & P (WA) & \cite{Diez-Alonso2019} & 4.67 (0.52) & +0.47 & \\
GJ 9689 & 3880 & -0.11 & 39.3 (0.4)  & S (HA) & \cite{Maldonado2021} & 7.35 (0.82) & +0.18 & \\
Gl 49 & 3740 & +0.13 & 18.4 (0.7)  & S (HA) & \cite{Suarez-Mascareno2018} & 1.55 (0.41) & +0.07 & binary\\
\hline
\end{tabular}
\begin{flushleft}
$^{1}$The full table is available as a machine-readable table (DOI:10.5281/zenodo.7578269)\\
$^{2}$P = photometric. S = spectroscopic.\\
$^{3}$AN=All-Sky Automated Survey for Super-Novae \citep[ASAS-SN,][]{Kochanek2017}, AP=APACHE \citep{Sozzetti2013}; AS=All Sky Automated Survey \citep[ASAS,][]{Pojmanski2002}; CA=CARMENES \citep{Quirrenbach2016}, ES=ESPRESSO \citep{Pepe2021}; FA=Fairborn \citep{Henry1999}; HA=HARPS \citep{Pepe2000}; HN=Hungarian Automated Telescope Network \citep[HAT-Net,][]{Bakos2004}; K2=\ktwo\ \citep{Howell2014}, KE=\kepler\ \citep{Borucki2010}; LC=Las Cumbres Observatory Global Telescope \citep[LCO,][]{Brown2013}; ME=MEarth \citep{Berta2012}, NS=Northern Sky Variability Survey \citep[NSVS,][]{Wozniak2004}; OS=Observatorio de Sierra Nevada; SP=SPIRou \citep{Donati2020}; ST=STELLA \citep{Strassmeier2004}; TE=\tess\ \citep{Ricker2014}; TJ=Telescope Joan Or\'{o} \citep[TJO,][]{Colome2010}; WA=Wide Angle Search for Planets \citep[WASP,][]{Pollacco2006}.\\
$^{4}$Metallicity-dependent age correction to be added to value for solar-metallicity.\\
$^{5}$If no period error is provided, an uncertainty of 1 day is assumed.
\end{flushleft}
\end{table*}

\emph{Binaries:}  Twenty-six of the 249 planet host stars are known to have stellar companions.  This 11\% fraction is much lower than 26.8$\pm$1.4\% among field M dwarfs \citep{Winters2019}.  Since exoplanet hosts are comparatively well-studied among cool field stars, this is very unlikely due to limited characterization of these stars.  Instead, it probably reflects survey/detection bias where binary stars are avoided in exoplanet surveys because it is usually more difficult to detect planets around them \citep{Kraus2016,Ziegler2018,Ziegler2021,Su2021,Clark2022}, and because contamination of the host star signal by other stars is detrimental to precise measurement of RVs \citep{Cunha2013}.

The rotational history of stars in multiple systems can differ greatly from that of single stars.  Stars with stellar companions are more likely to be rapidly rotating relative to single stars of the same age/mass \citep{Kraus2012a,Simonian2019}.  Very close ($\ll 1$ au) binaries transfer orbital AM to rotational AM via tidal torques \citep{FlemingD2019}.  At moderate separations ($\sim$100 au) a stellar companion will truncate a disk and shorten its viscous lifetime \citep{Cieza2009,Kraus2012a,Rosotti2018}.  This removes a sink of stellar angular momentum, allowing the star to spin-up unimpeeded during pre-main sequence contraction \citep{Messina2019}.  

\begin{figure}
    \centering
    \includegraphics[width=\columnwidth]{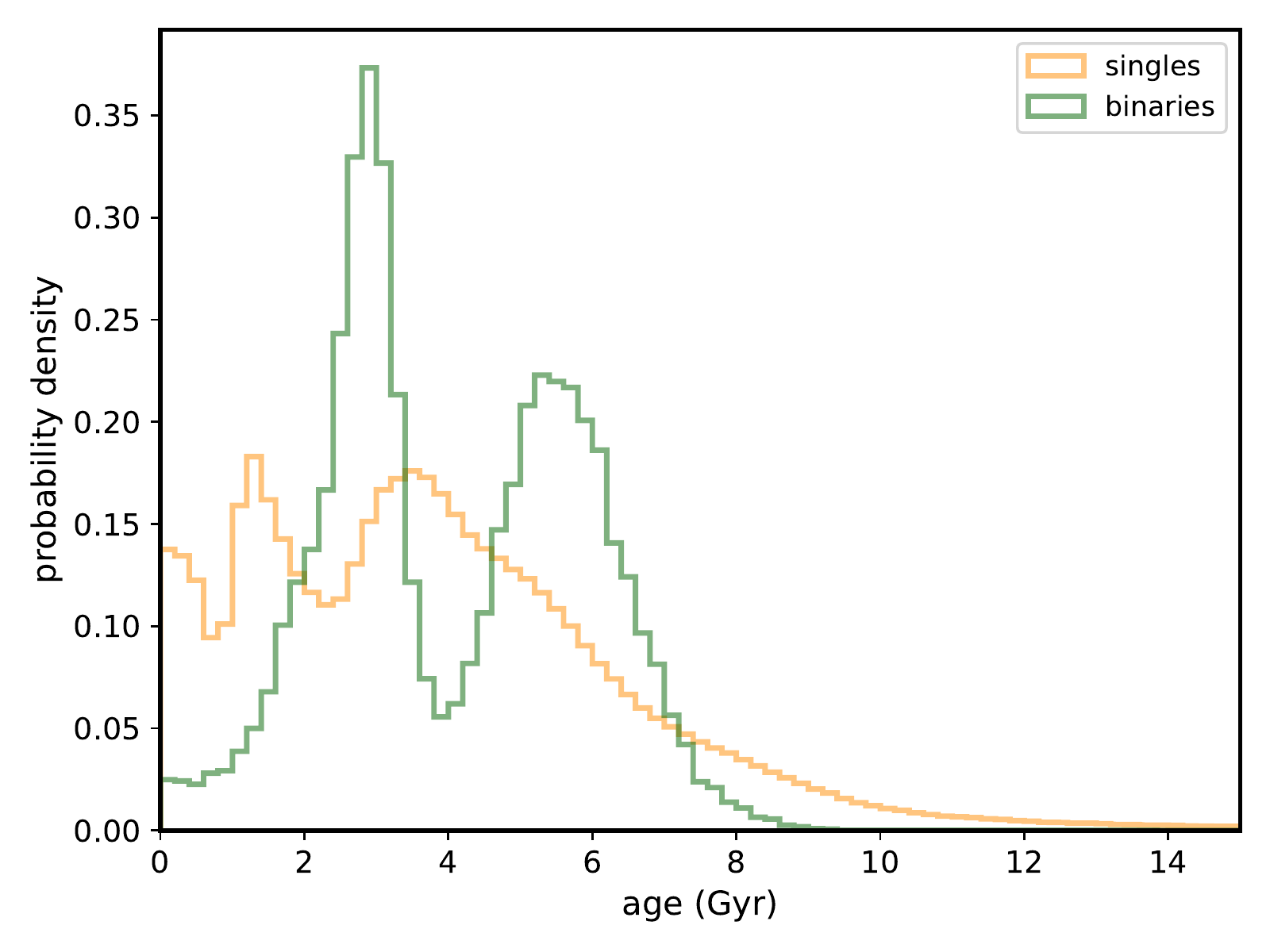}
    \caption{Distribution of estimated ages for single vs. binary/multiple host stars, accounting for the uncertainties.  The irregularity of the binary distribution is sampling noise: only 26 systems constitute the binary sample.}
    \label{fig:binaries}
\end{figure}

We computed ages based on rotation without regard to multiplicity, but warn that in the case of binaries, such ages could be seriously in error.  In this sample, at least, the distribution of rotation periods does not appear remarkably different from single stars (Fig. \ref{fig:periods}) nor does the age distribution of known binaries appear remarkable (Fig. \ref{fig:binaries}), although it is greatly limited by the small sample size.  This could be because the smaller fraction of binaries that do appear in the catalog tend to be very wide, and the effects on rotation and hence age are negligible.   

\emph{Anomalously old stars:} Four host stars (GJ 27.1, GJ 667 C, HD 238090, and HIP 70849) are assigned problematic ages that are $>2\sigma$ older than 10 Gyr, the nominal age of the Galactic disk.  These are unlikely to be Galactic halo or former globular cluster members because unusual abundances and peculiar motion characteristic of such stars would have been noted.  Three of the stars are in binaries, in which stellar companions could directly or indirectly affect the rotation evolution.  The fourth, GJ 27.1, has a 140$\pm$10 day rotation period estimated from ASAS-SN photometry, far longer than that expected for an early-type M dwarf in the Galactic disk, and the age is obviously unphysical: $35 \pm 9$ Gyr.  The star's metallicity has not been reported.  Potentially the rotation period is an artifact, e.g., confusion with another star (the survey's resolution is 15").

\emph{Young stars:} No PMS-ages were assigned to our stars, which is expected since we removed all known disk-hosting and PMS stars from our catalog.  Fourteen stars can only be assigned upper limits for ages, and of these 8 are younger than 1 Gyr at the 95-percentile level and not known to be binaries.   The rotation period of one of these, TOI-620, is tentative and the star is also a suspected binary \citep{Reefe2022}.  The other seven are \ktwo-detected systems: K2-43, K2-239, K2-240, which hosts two transiting Neptune-size planets, has been detected in X-rays \citep{Foster2022}; K2-284, previously reported having a young age by \citet{David2018}, K2-324, K2-354, and KOI-5879, a flaring M dwarf \citep{Yang2019}. 

\subsection{Individual Noteworthy Systems}
\label{sec:individual}

\emph{GJ 229:} The nearby (5.76 pc) M1 dwarf Gliese/GJ 229 has an ultra-cool (T7-type) dwarf companion \citep{Nakajima1995}.  The primary's 27.3-day rotation period was determined from ASAS-SN photometry \citep{Suarez-Mascareno2016} and we estimate an age of $3.8 \pm 0.5$ Gyr, with the caveat that the existence of the companion on a 29 au orbit could have affected the rotation history.  No uncertainty in \prot\ was reported but this is likely to be small given the multi-year baseline of ASAS-SN.

\emph{GJ 1214:} This nearby M4-type dwarf with a well-studied transiting ``sub"-Neptune-size planet on a 1.58-day orbit \citep{Charbonneau2009}.  Based on the $125\pm5$ day rotation period identified in STELLA photometry by \citet{Mallonn2018}, we estimate an age of $5.9 \pm 3.1$~Gyr.   

\emph{LHS-1815:} This M1-type dwarf (aka TOI-704) hosts a transiting Earth-size planet on a 3.8-day orbit.  The star lies 1.8 kpc above the Galactic plane and kinematically belongs to the ``thick" Galactic disk population \citep{Gan2020}.  We estimate an age of $7.3 \pm 1.3$ Gyr, consistent with the expected age of that population.   

\emph{K2-22:} K2-22 is a late K dwarf that hosts what has been proposed to be a ``evaporating" planet on a 9-hour orbit that manifests itself as quasi-periodic dimming due to accompanying dust cloud \citep{Sanchis-Ojeda2015}.  The highest peak in a Lomb-Scargle periodogram in \ktwo\ photometry; is at $7.61\pm0.26$ days but the shape of the lightcurve (Fig. \ref{fig:ktwo1}) suggests this is one-half the period \citep{Sanchis-Ojeda2015}.  A period of $15.2 \pm 0.5$ days yields an estimated age of $1.1 \pm 0.2$ Gyr, but this star has an M dwarf companion at a projected separation of 460 au \cite{Sanchis-Ojeda2015} which could have affected its rotation history.  

\emph{Barnard's Star (GJ 699):} This very metal-poor, high peculiar motion M4-type is classified as intermediate between the Galactic Disk and Halo populations \citep{Gizis1997}; a putative Doppler RV-detected planet \citep{Ribas2018} around this star has been disputed \citep{Lubin2021}. It is \emph{not} in our current catalog because its \teff\ is marginally cooler than our 3200K cut-off, but, motivated by its unusual nature and recently confirmed \prot\ of 145$\pm$15 days \citep{Toledo-Padron2019,Terrien2022b}, we compare this to the cool extremum of the \citet{Dungee2022} M67 gyrochrone, which reaches 120 days at 3250K (Fig. \ref{fig:gyro_mc}) and at cooler temperatures is essentially an unconstrained extrapolation.  Thus the gyrochronology suggests an age older than M67, as expected, but extension of M dwarf gyrochrones into the fully convective area is needed before assigning any robust age.

\section{Summary and Discussion}
\label{sec:lastsection}

\subsection{Rotation and Ages of M dwarf Exoplanet Hosts}
\label{sec:summary}

\begin{figure}
    \centering
    \includegraphics[width=\columnwidth]{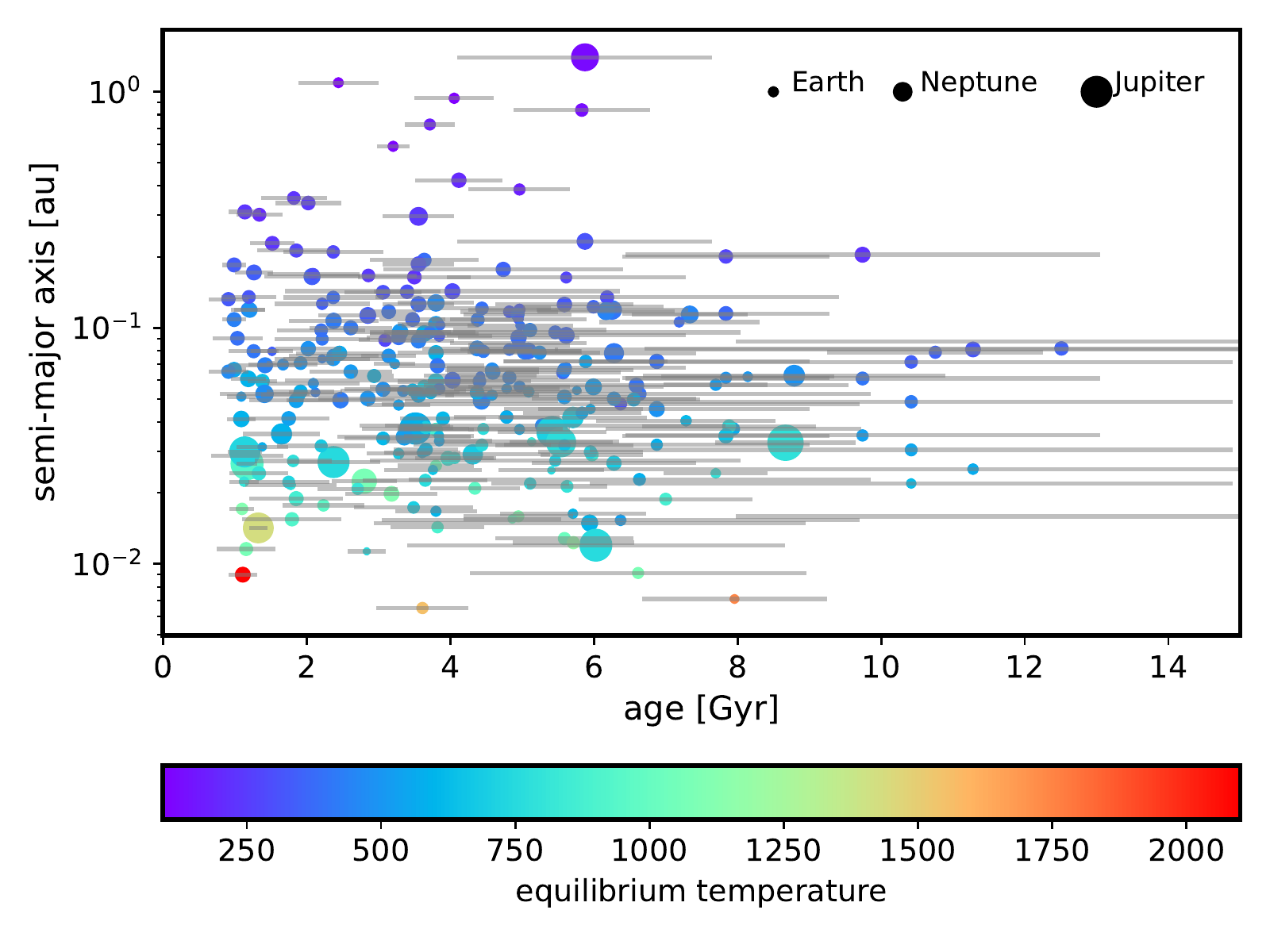}
    \caption{111 confirmed planets with properties from the NASA Exoplanet Archive with age estimates from this work.  Points are scaled with planet radius and color-coded by planet equilibrium temperature.  Ages have not been filtered for binarity nor corrected for metallicity.}
    \label{fig:planets}
\end{figure}

The value of robust ages for exoplanet studies, and advances in the gyrochronology of older cool M dwarfs motivated us to catalog rotation periods among late K and early M-type dwarfs (\teff=3200-4200K) that host known planets, and to apply empirical, \teff-dependent rotation-age relations to estimate ages and their standard errors.  This complements work on calibrated rotation-based ages among younger PMS stars \citep{Kounkel2022}.  We cataloged 249 stars with rotation periods, 227 of which we are able to estimate ages with a median error of 20\% and mode of 14\%, and to an additional 8 we assign upper limits (Table \ref{tab:time}).  Our fractional error is significantly higher than the 5-10\% estimated by \citet{Otani2022}, probably because we include additional potential sources of error.  Figure \ref{fig:planets} shows ages of candidate or confirmed planets around these stars and the distribution with semi-major axis, radius, and equilibrium temperature, as reported in the NASA Exoplanet Archive.

The age distributions of both \kepler\ and non-\kepler\ host stars peak at around 3 Gyr with a steady decline to near zero at 10 Gyr, the age of the Galactic disk (Fig. \ref{fig:age_dist}).  The resemblance of the actual and ``mock" populations (the latter with a uniform 0-10 Gyr age distribution) shown in Figure \ref{fig:mock} indicates that the structure of the distribution, particularly at young ages, is partly due to the discontinuous and limited coverage of the current gyrochronology and dispersion of the distribution due to error.  The peaks at $<$3 Gyr correspond approximately with the location of the calibration ages, and are likely artefacts due to ages that cannot be assigned in certain regions of \teff-\prot\ space and have only upper limits assigned.  Other effects impacting the derived distribution include the opposing biases  against detection of planets around younger, more rapidly rotating, and more active stars \citep{Miyakawa2022}, and against detection of rotational variability among older, slowly rotating, less active stars \citep{Morris2020}.  \footnote{``Stalling" of spin-down due to core-envelope decoupling would result in \emph{broadening} of the age distribution, not peak formation.}  Several versions of the local star formation history based on white dwarf cooling ages and \gaia\ astrometry also peak at around 3-5 Gyr \citep{Isern2019,Mor2019,Alzate2021}, so the distribution of host star ages could reflect this. 
Our error analysis (Sec. \ref{sec:error}) shows that one limiting source of error could prove to be the precision of stellar parameters, i.e. \teff\ and [Fe/H].  The sensitivity to \teff\ is due to the steepness of rotation sequences for very cool dwarfs.  \teff, which is related to the surface brightness and hence convective vigor of a star, is the appropriate independent variable for gyrochronology, but must be inferred from observables.  The ultimate limit on accurate values of \teff\ precision for M dwarfs is the challenge of establishing a reliable temperature scale.  
\subsection{Caveats and Limitations}
\label{sec:caveats}

We have adopted the values and standard errors of \prot\ from the literature at face value.  The possibility that the true period is twice the published value needs to be considered in cases of stars with anomalous rapid rotation and young ages.  Ground-based observations can also suffer from aliasing imposed by diurnal, lunar, and annual window functions.  

Our gyrochronology assumes that the narrow rotational sequence observed among the late K dwarfs and the warmer M dwarfs in the 2.7 Gyr-old Ruprecht 147 cluster \citep{Curtis2020} extends to 3200K by 2.7 Gyr, and that the $n=0.62$ power-law spin-down derived by \citep{Dungee2022} also applies to cooler M dwarfs at later times.  This assumption could fail if the rotational sequence among M67 M dwarfs was formed by stalling, rather than a transition from saturated to un-saturated braking laws.  Core-envelope re-coupling is expected to become weaker and take longer towards the fully convective boundary, which could mean that a rotational sequence appears much later, or not at all.  This depends in part on the \teff\ or mass dependence of the core-envelope coupling time vs. $\tau_c$.  This issue can only be resolved by deeper  monitoring of Ruprecht 147 or a cluster of similar age.

Another contributor to our error budget is sensitivity of the rotation-age relation to non-solar [Fe/H]. In the absence of appropriate calibration, we relied entirely on theoretical models to estimate this effect.  The part due to changes in the structure and moment of inertia of the star is reasonably well-constrained by observations, but magnetic field pressure could modulate this.  A model that includes this effect \citep{Feiden2016} predicts that at a fixed \teff=3700K, inclusion of magnetic pressure increases the moment of inertia by 40\%, but this is almost entirely due to a change in the mass inferred for a given \teff.      More uncertain is the metallicity-dependence of the torque, which, at least in our models, dominates the sensitivity.  We based this scaling on the magnetized wind formulation of \citet{Matt2012} as reformulated by \citet{vanSaders2013}, but this was developed for solar-type stars with rotationally-aligned dipole fields.  

Finally, rotation-based ages for binary systems must be carefully considered, particularly given the possibility of a third, closer and unresolved component \citep{Reipurth2012}.  While most published exoplanets have had some sort of screening for binaries, not all of them cover all the parameter space, and surveys of very cool KOIs are only now coming to fruition.

\subsection{Outlook}
\label{sec:outlook}

Since \teff\ is not an observable and cannot be readily derived without stellar radii, gyrochrones could be established in a common reddening-corrected color which is also available for stars of interest; ideally, this color should be directly related to \teff, it should be relatively [Fe/H]-independent.  It should also use redder filters in which M dwarfs are comparatively bright and reddening is smaller.  These requirements impact the use of \gaia\ photometry since M dwarfs are faint in the $B_p$ synthetic band used to construct $B_p-R_p$ colors.  Analysis of color magnitude diagrams of \kepler\ M dwarfs using PanSTARRS photometry suggest that $g$-$Y$ holds promise (A. Ali, pers. comm.).  

Establishing precise M dwarf metallicities is a work in progress \citep[e.g.,][]{Passegger2021}.  More challenging will be to validate the effect of metallicity on rotation-age relations, which here we have here treated only via stellar interior models and torque-law scaling.   Tests of the metallicity-scaling of the torque law are desperately needed. Metal-poor or metal-rich clusters are relative rare \citep{Heiter2014} and thus, statistically, found at greater distances where observations to establish rotational sequences will be challenging.  The wealth of binaries provided by \gaia\ \citep{El-Badry2021} might serve as a road to calibration over a wider range of stellar parameters \citep{Otani2022}, provided sufficiently wide examples can be identified and precise ages for the primaries can be determined by other means.

Gyrochronology of exoplanet host stars is an ongoing effort and we envision the TIME-Table to be a ``living" catalog of very cool dwarf rotation periods and ages that is periodically updated and revised with new discoveries and advances in gyrochronology.  New planetary systems are constantly being detected, validated, or confirmed, particularly by the \tess\ mission, now in its fifth year.  Rotational variability is being detected by two ongoing space surveys: \tess\ and, with much longer baseline but much sparser cadence, \gaia\ \citep{Distefano2022}.  Ground-based surveys like ZTF and the Rubin Observatory \citep{Hambleton2022} can provide observations of distant field stars and young clusters.  Although the \tess\ 27-day sector interval severely limits its ability to detect the rotation of older field stars \citep{Claytor2022}, those stars in and around the two Continuous Viewing Zones around the ecliptic poles are observed for multiple sectors and in principle it is possible to detect longer periods \citep{Hedges2020b,Claytor2022}.

Many rotation periods have been established by analysis of Doppler RV residuals or indicators of activity in the time-series high-resolution spectroscopy obtained to detect, confirm, or measure the masses of planets \citep{Suarez-Mascareno2015}.  \citet{Terrien2022b} showed that detection of the periodic signal in the Zeeman broadening of lines can reveal rotation of magnetic active regions on the star and yield a rotation period.  The Zeeman effect increases with wavelength-squared, and the proliferation of high-resolution spectrographs operating in the infrared could lead to additional rotation periods using this approach.  Alternatives to Lomb-Scargle periodogram analysis which are more robust to spot evolution such as autocorrelation and Gaussian process regression \citep{Angus2017,Nicholson2022} could be used to obtain more precise ages.  Age-dating could also adopt a Bayesian approach, with Galactic population age distributions as priors \citep[e.g.,][]{Mor2019,Cukanovaite2022}.

Last but not least, additional observations of open clusters for calibration will improve the gyrochronology and lead to more precise (and hopefully more accurate) ages.  In particular, the \teff\ range of existing gyrochrones (including M67) should be extended through the fully convective boundary to include mid- and late-type M dwarfs representing hosts stars of particular interest (e.g., TRAPPIST-1), and, foremost, to establish whether a narrow rotational sequence appears by a few Gyr --- without which gyrochronology is futile.  The small effective area and large pixel size of \tess\ greatly limits its utility here, since older clusters are rare and hence more distant.  The \emph{Plato} mission will offer only limited improvement over \tess\ (15" vs. 20" pixels).  However, the \emph{Roman Space Telescope} will have a field of view of 0.28 $\mathrm{deg}^2$ with 0.11" pixels.  Ages of calibrator clusters could see refinement from a combination of \gaia\ parallaxes, asteroseismology, and high-throughput spectroscopy \citep{Fu2022}.  Otherwise, much of the observations need to be performed from the ground using wide-field telescopes with sufficient aperture.  Wide-field adaptive optics can alleviate the issues of source confusion in the fields of more distant clusters. For example, ground-layer adaptive optics (GLAO) can provide a factor of 2--3 improvement in spatial resolution compared to seeing-limited observations while still capturing an entire cluster in one observation \citep{Rigaut2002}.  Such improvements enable observations of dwarfs to spectral type M7 in the majority of clusters older than 1 Gyr \citep{dungeethesis}\footnote{https://scholarspace.manoa.hawaii.edu/collections/67c8d51f-97a4-4d45-8a69-8baeaacaeaf6}.

\section*{Acknowledgements}

This work is dedicated to the memory of F.C.G., for whom the very stars of heaven were new.  E.G. and R.D. were supported by NSF Astronomy \& Astrophysics Research Program Grant 1817215.  We thank Jen van Saders and an anonymous reviewer for helpful feedback on earlier versions of this manuscript.  This paper includes data collected by the \kepler\ and \tess\ missions and obtained from the MAST data archive at the Space Telescope Science Institute (STScI). Funding for the Kepler mission is provided by the NASA Science Mission Directorate. STScI is operated by the Association of Universities for Research in Astronomy, Inc., under NASA contract NAS 5–26555.  This paper makes use of data from the MEarth Project, which is a collaboration between Harvard University and the Smithsonian Astrophysical Observatory. The MEarth Project acknowledges funding from the David and Lucile Packard Fellowship for Science and Engineering and the National Science Foundation under grants AST-0807690, AST-1109468, and AST-1004488 (Alan T. Waterman Award), and a grant from the John Templeton Foundation.

\section*{Data Availability}

All data used in this work are in the public domain and available through various archives.




\appendix

\section{Lightcurve Analysis}

\begin{figure*}
\centering
\includegraphics[width=\textwidth]{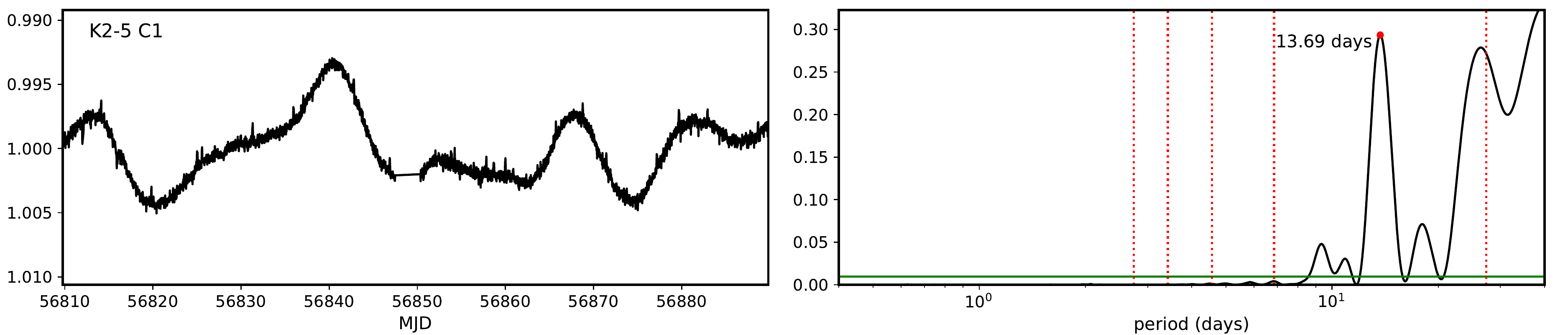}
\includegraphics[width=\textwidth]{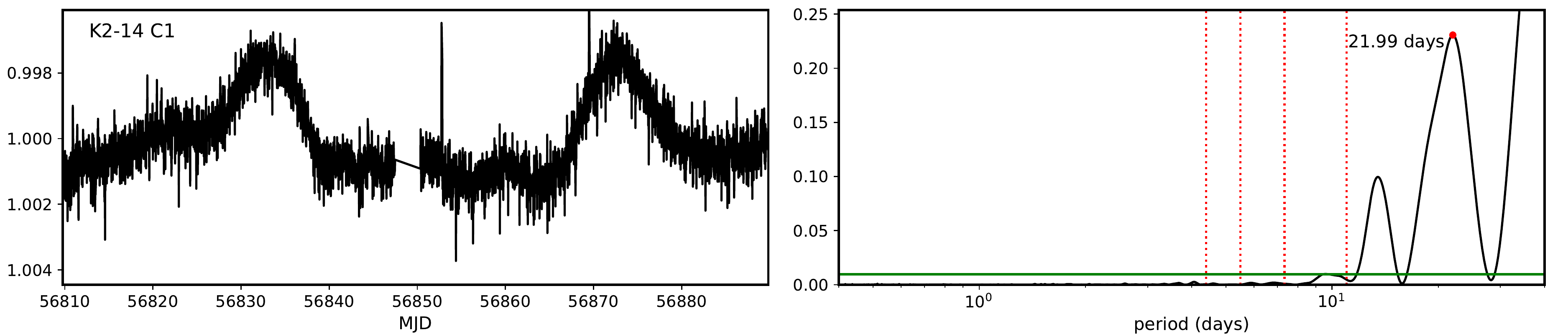}
\includegraphics[width=\textwidth]{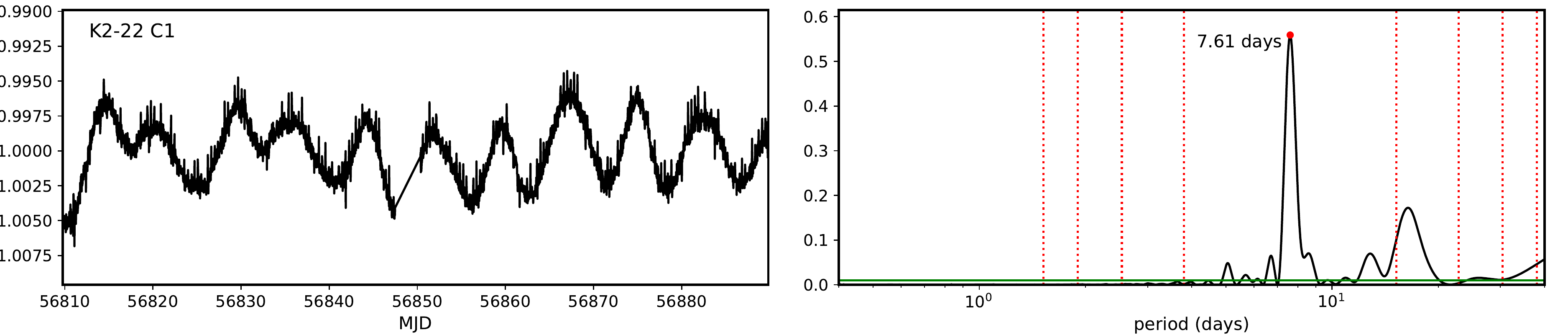}
\includegraphics[width=\textwidth]{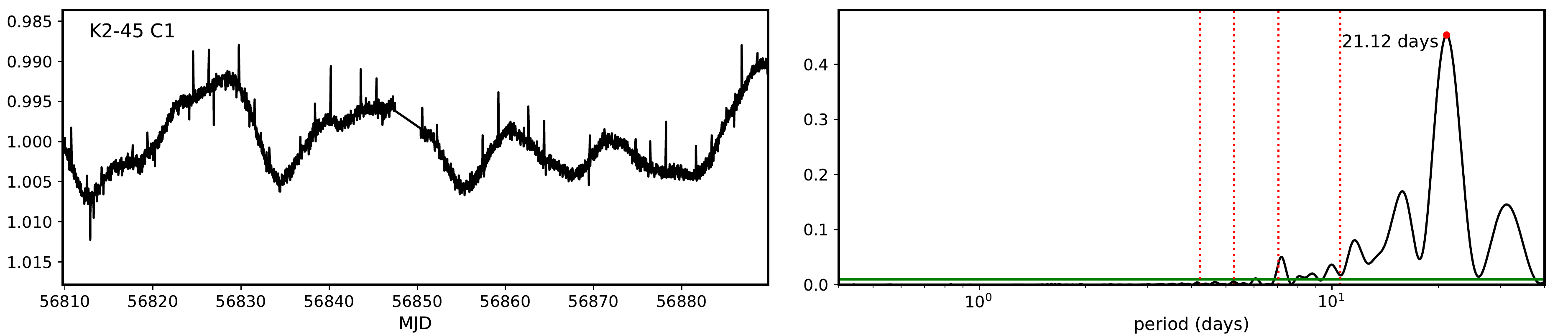}
\includegraphics[width=\textwidth]{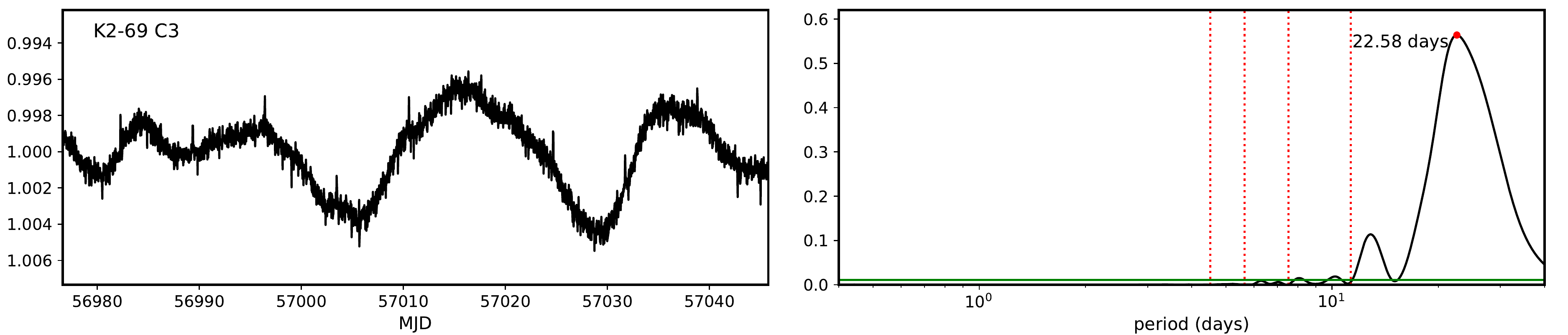}
\includegraphics[width=\textwidth]{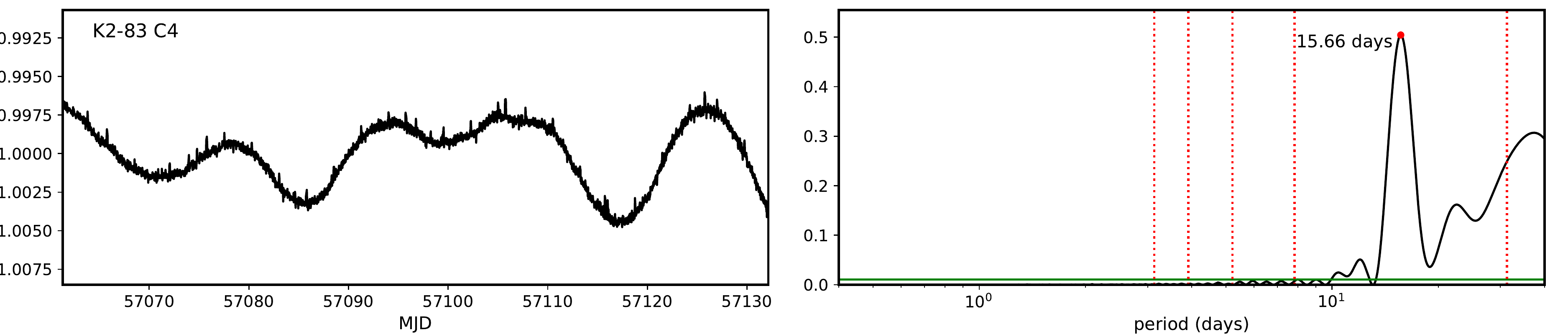}
\vspace{-0.1in}
\caption{De-trended \ktwo\ PDCSAP lightcurves of M dwarf host stars with significant periodic signals.  The left panels contain the lightcurves.  The right panels show the Lomb-Scargle periodograms with the horizontal green line marking the false alarm probability $p=0.001$, the red point marking the period of highest power, and the vertical dashed lines marking upper and lower harmonics of that period.  For K2-22, twice the peak period was adopted.}
\label{fig:ktwo1}
\end{figure*}

\begin{figure*}
\centering
\includegraphics[width=\textwidth]{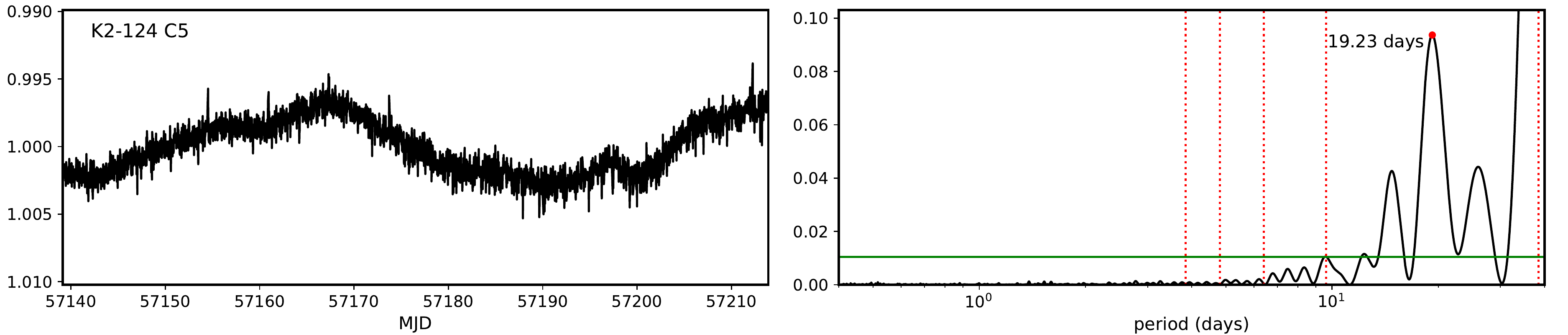}
\includegraphics[width=\textwidth]{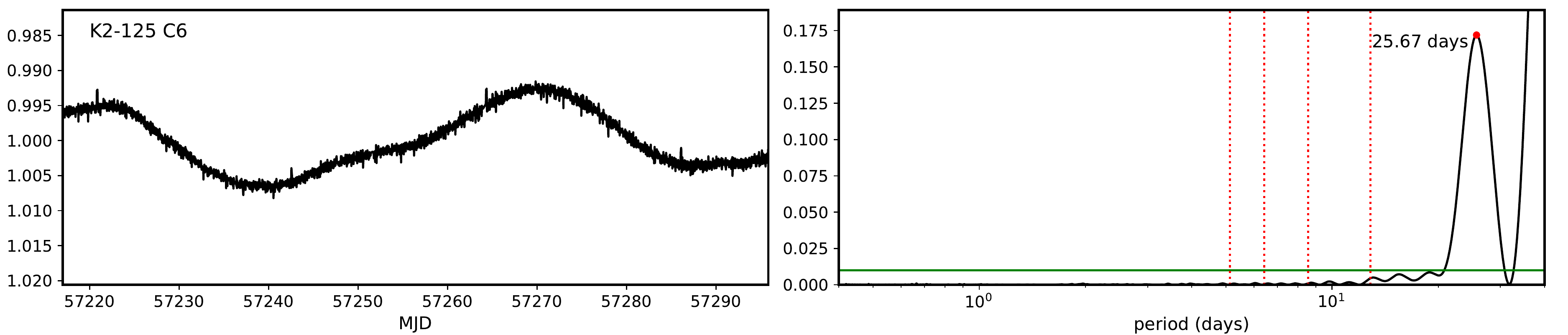}
\includegraphics[width=\textwidth]{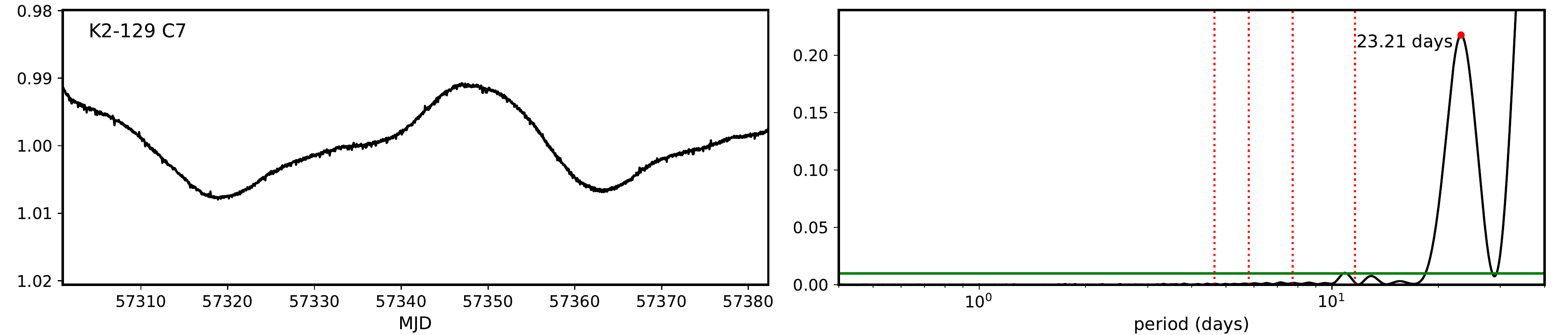}
\includegraphics[width=\textwidth]{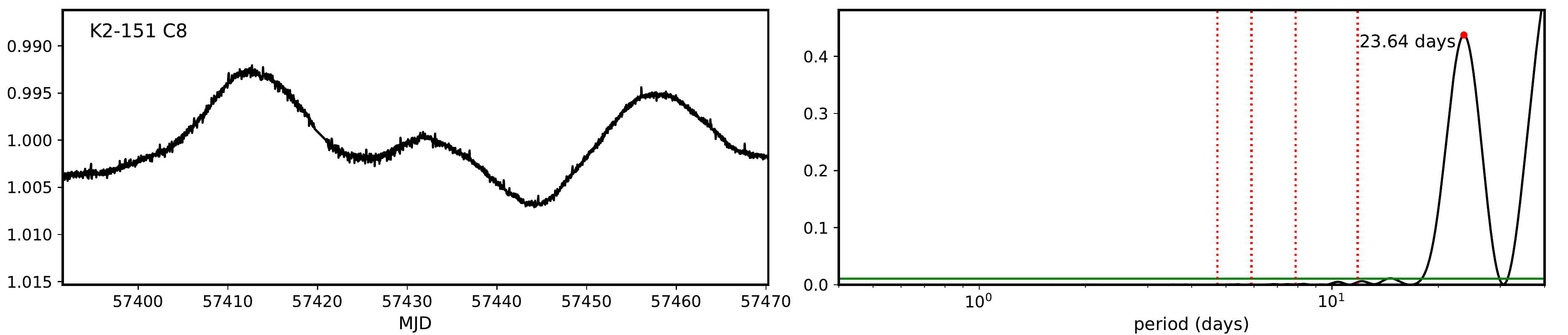}
\includegraphics[width=\textwidth]{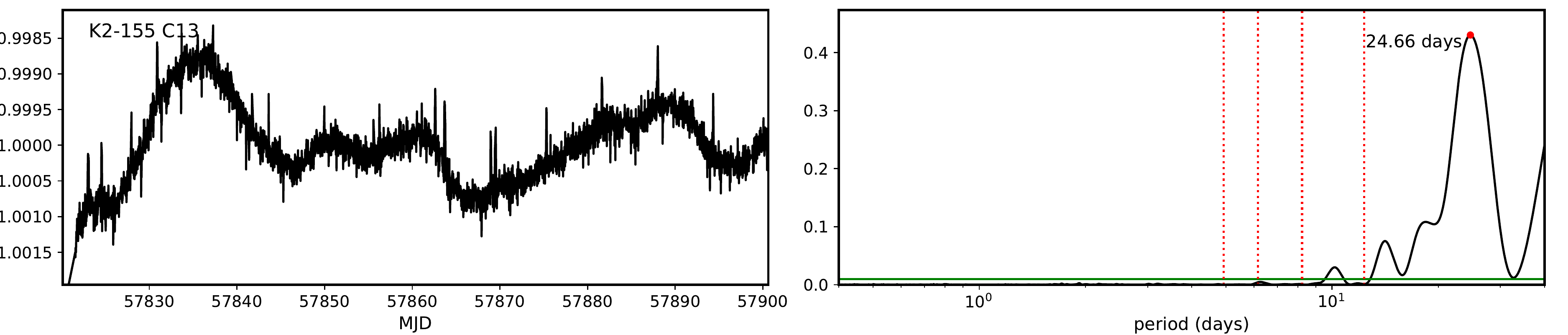}
\includegraphics[width=\textwidth]{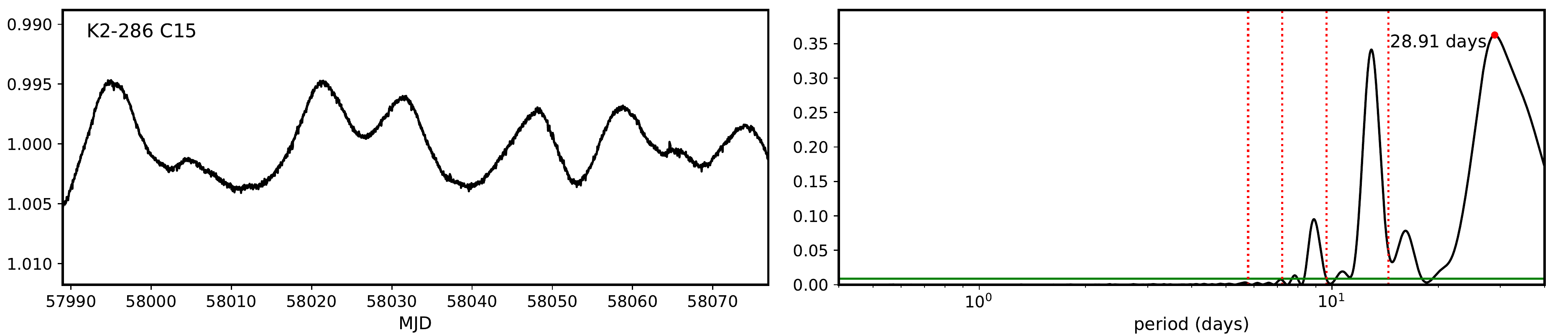}
\caption{Additional \ktwo\ lightcurves.  See Fig. \ref{fig:ktwo1} for explanation.}
\label{fig:ktwo2}
\end{figure*}

\begin{figure*}
\centering
\includegraphics[width=\textwidth]{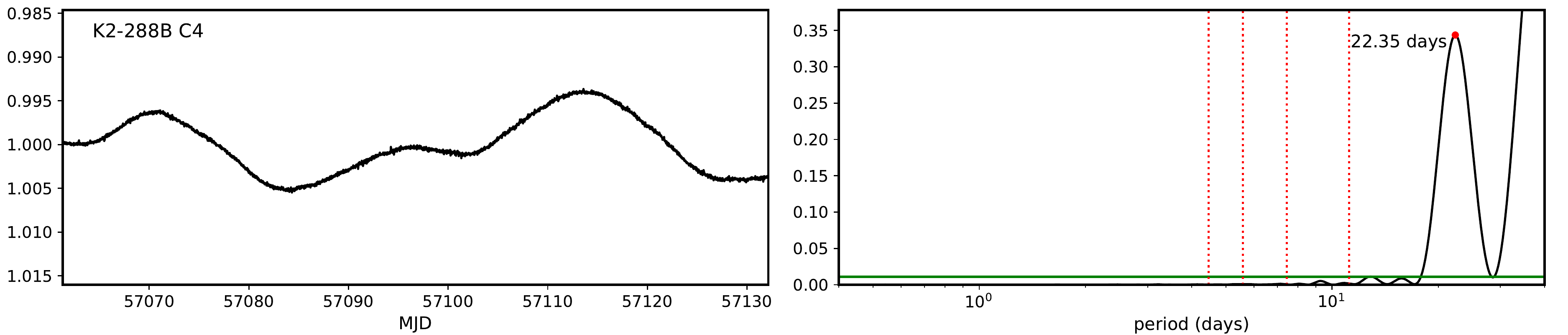}
\includegraphics[width=\textwidth]{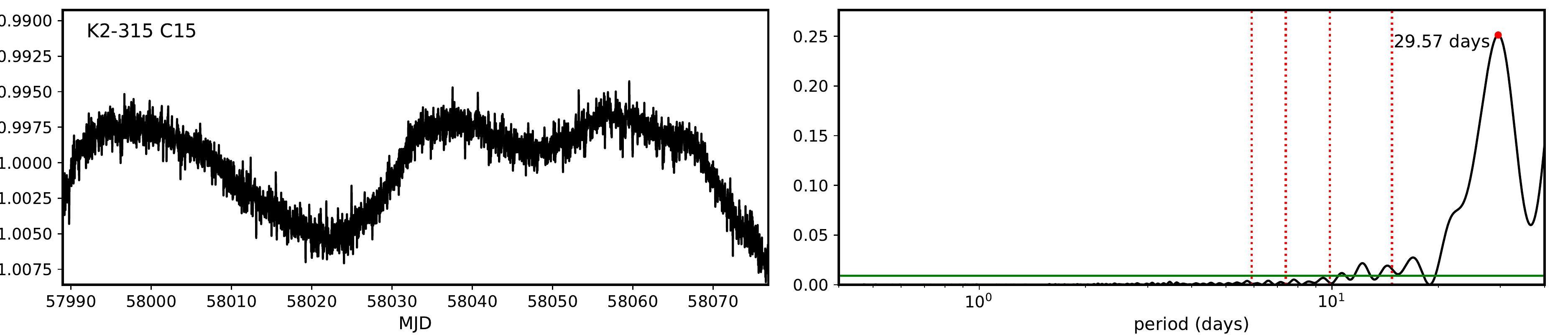}
\includegraphics[width=\textwidth]{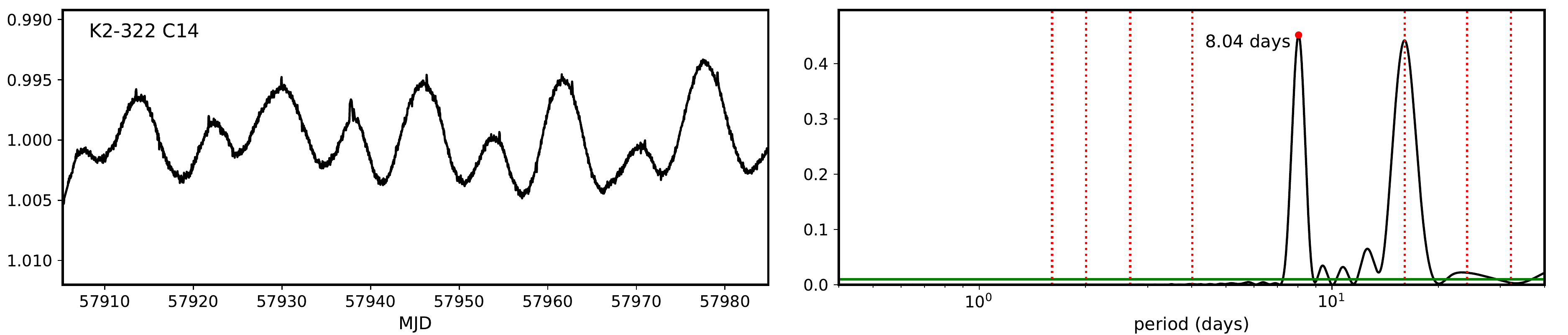}
\includegraphics[width=\textwidth]{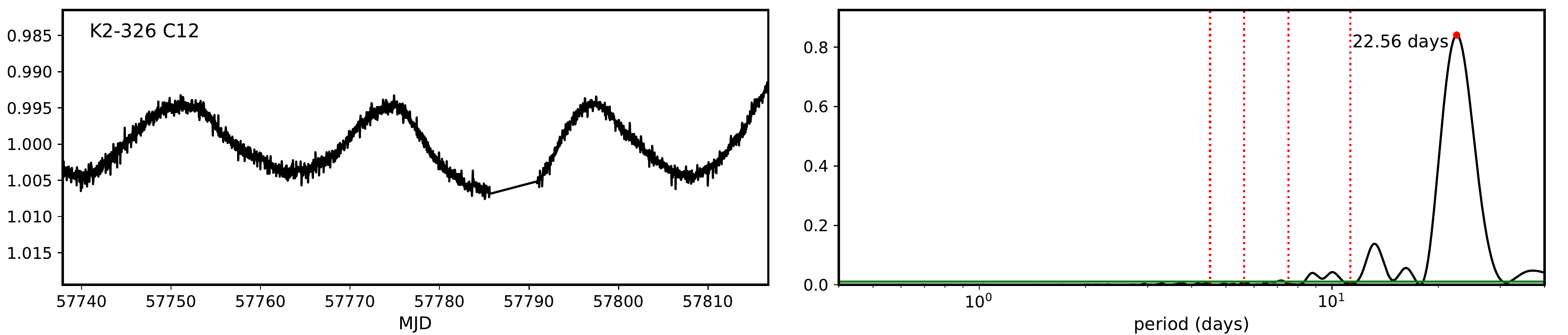}
\includegraphics[width=\textwidth]{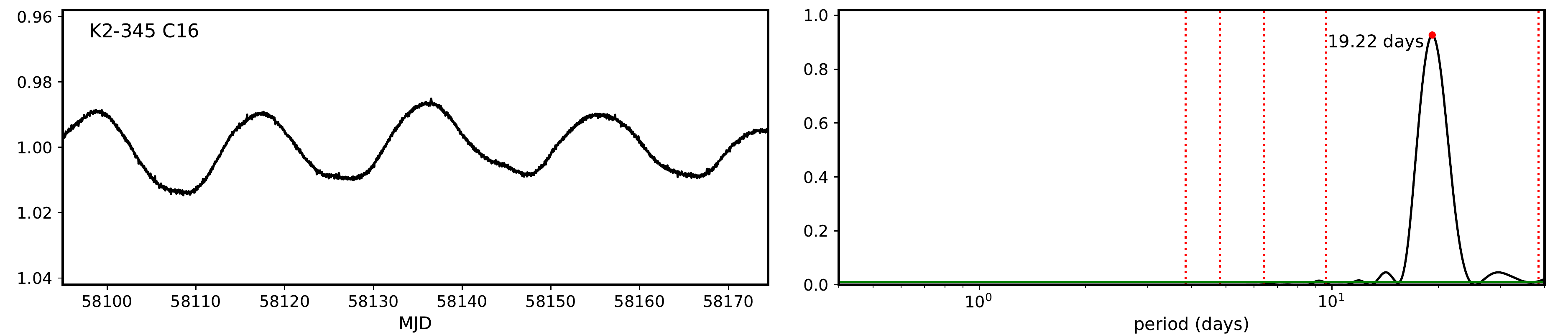}
\includegraphics[width=\textwidth]{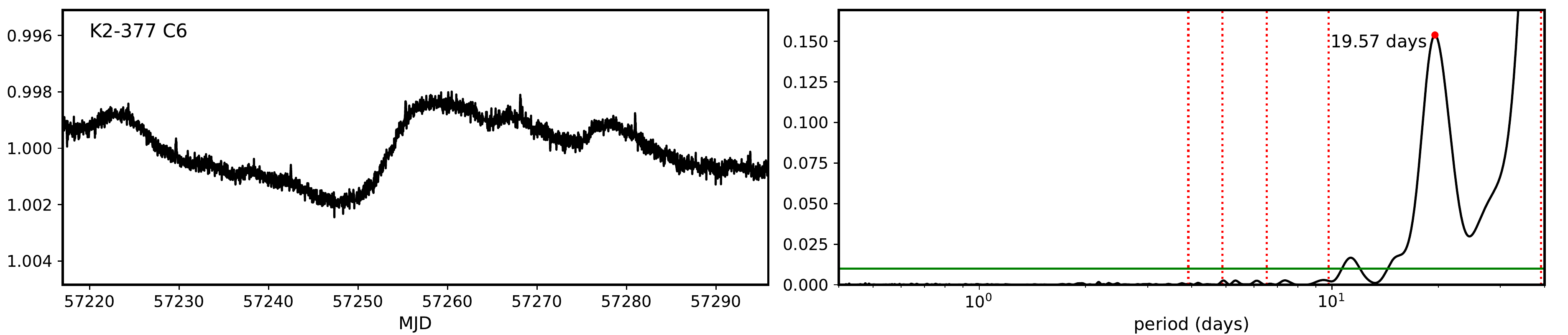}
\caption{Additional \ktwo\ lightcurves.  See Fig. \ref{fig:ktwo1} for explanation.  For K2-322, twice the peak period was adopted.}
\label{fig:ktwo3}
\end{figure*}

\begin{figure*}
\centering
\includegraphics[width=\textwidth]{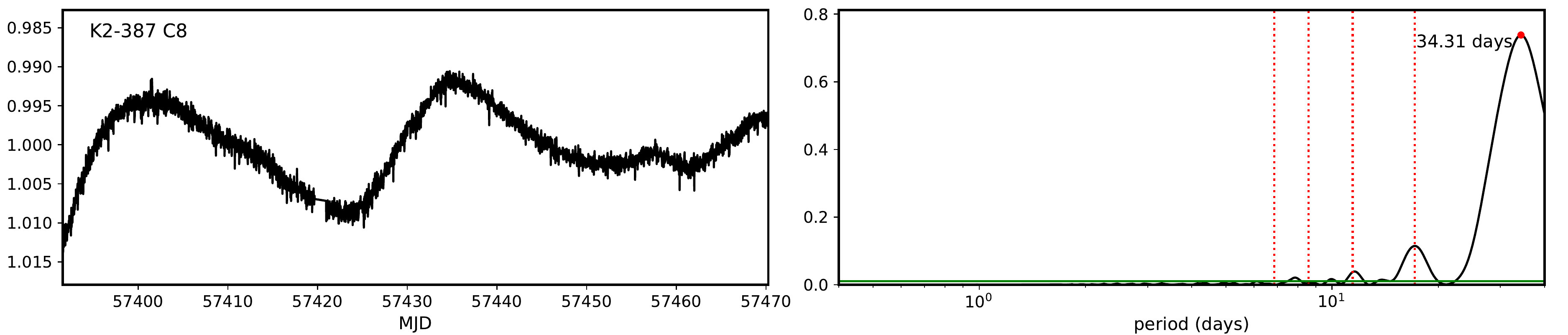}
\includegraphics[width=\textwidth]{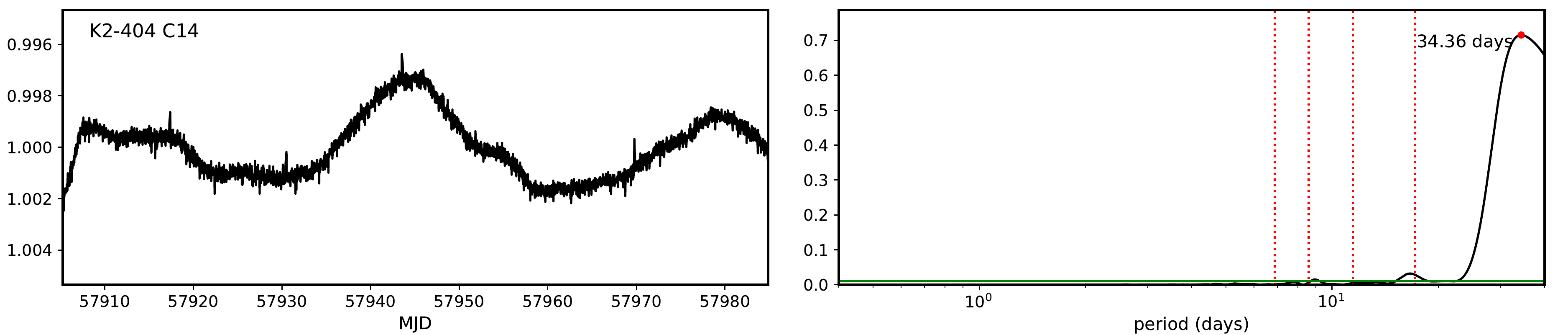}
\includegraphics[width=\textwidth]{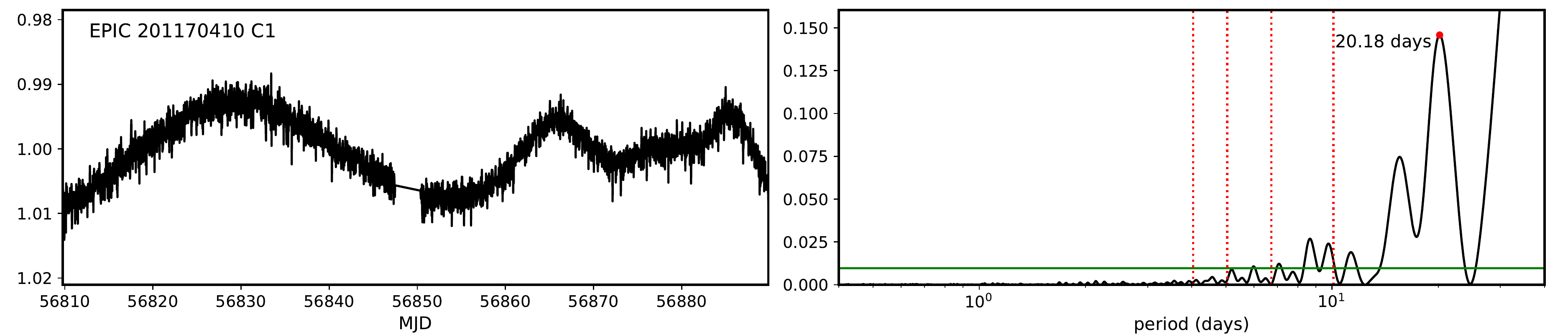}
\caption{Additional \ktwo\ lightcurves.  See Fig. \ref{fig:ktwo1} for explanation.}
\label{fig:ktwo4}
\end{figure*}

\begin{figure*}
\centering
\includegraphics[width=\textwidth]{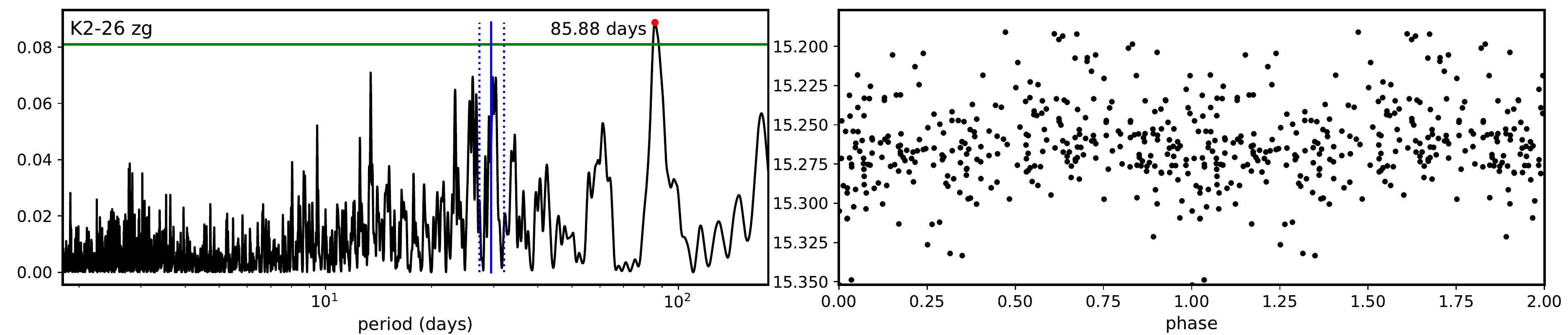}
\includegraphics[width=\textwidth]{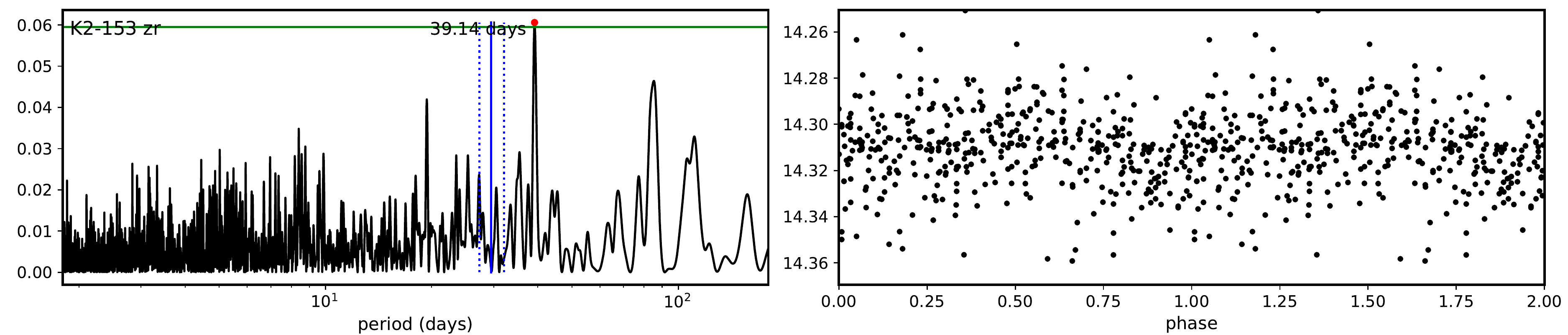}
\includegraphics[width=\textwidth]{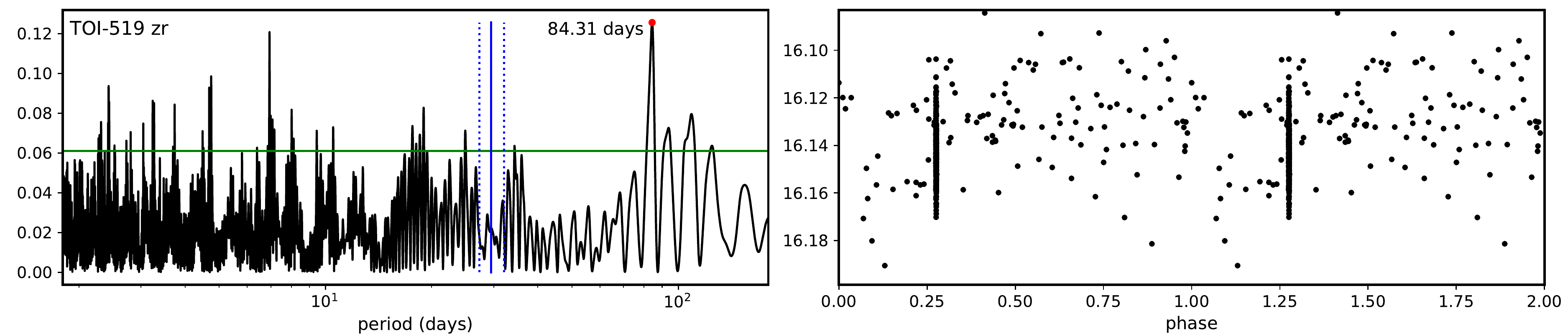}
\includegraphics[width=\textwidth]{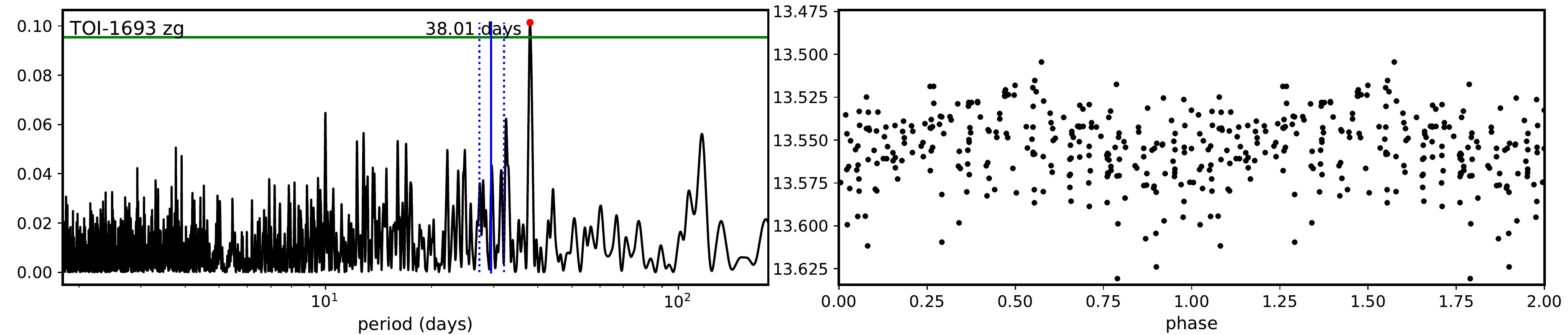}
\caption{Periodograms and phased ZTF lightcurves of four M dwarf exoplanet hosts with significant ($p<0.01$, horizontal green line) signals (red dots) that passed our visual inspection and are considered candidate rotational signals.  The host star and the band-pass are indicated in the The blue solid and dashed lines are the lunar synodic period and its aliases with the annual observing window function.  Note that the phased lightcurves are repeated.}
    \label{fig:ztf}
\end{figure*}

\begin{figure*}
\centering
\includegraphics[width=\textwidth]{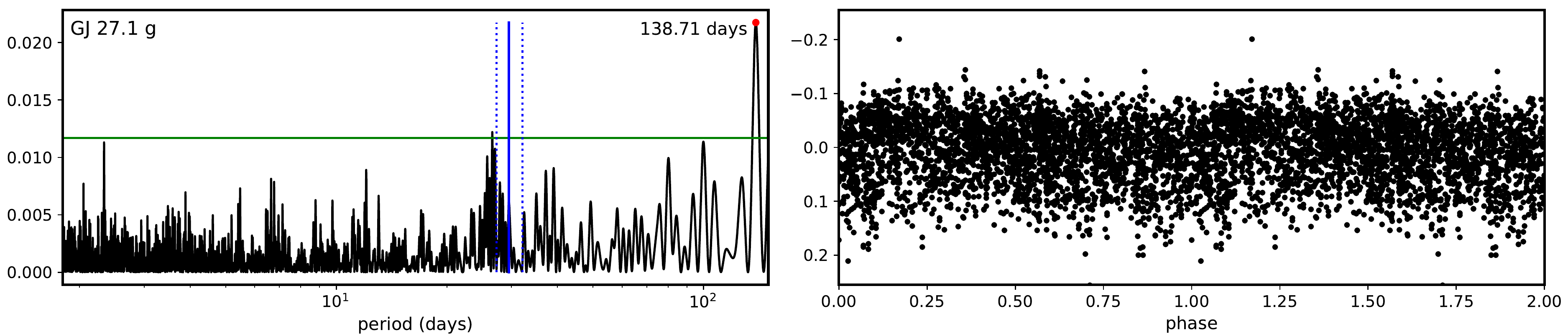}
\includegraphics[width=\textwidth]{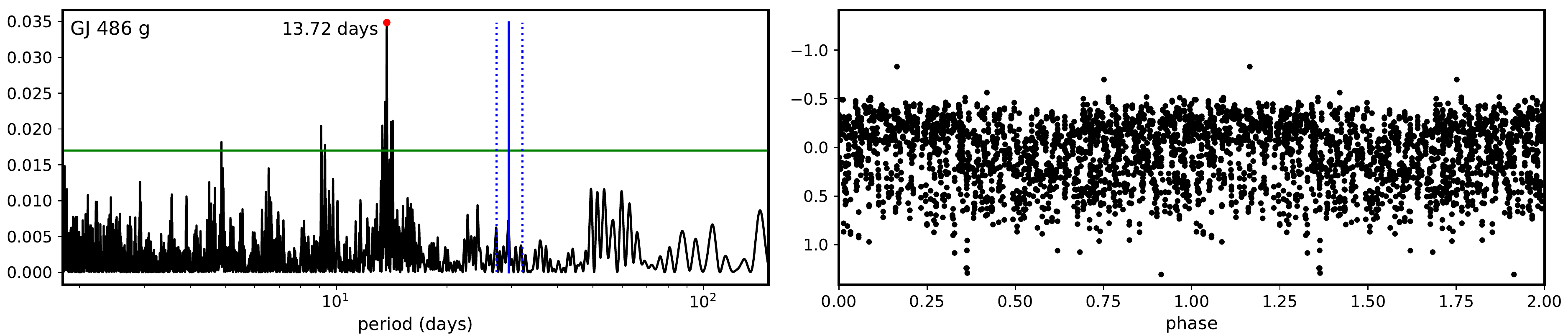}
\includegraphics[width=\textwidth]{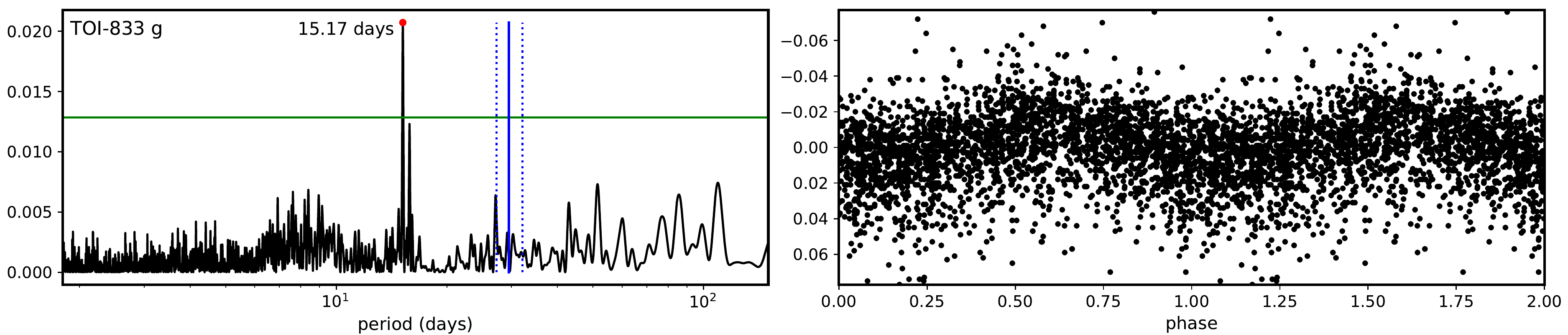}
\includegraphics[width=\textwidth]{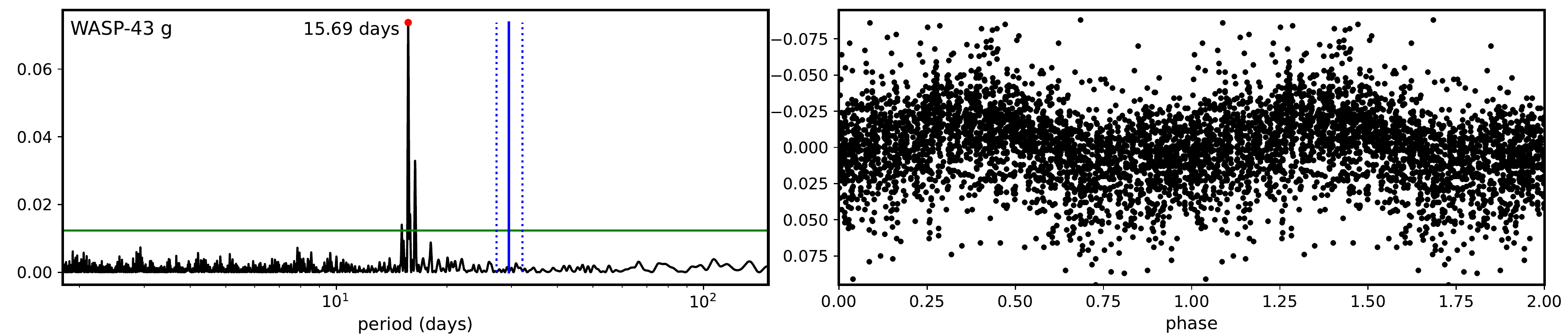}
\caption{Periodograms and phased ASAS-SN $g$-band lightcurves of four M dwarf exoplanet hosts with significant ($p<0.01$, horizontal green line) signals (red dots) that also had equivalent significant signals in $V$-band, passed our visual inspection, and are considered candidate rotational signals.  The blue solid and dashed lines are the lunar synodic period and its aliases with the annual observing window function.  Note that the phased lightcurves are repeated.}
    \label{fig:asas-sn}
\end{figure*}

Figures \ref{fig:ktwo1}-\ref{fig:ktwo4} show \ktwo\ lightcurves of 21 host star in which new rotational signals  were identified or, in the case of K2-345, replace a previously published value.  In the cases of K2-5, 14, 83, 124, 125, 129, 151, 288\,B, 315, 322, and 377, a rotation period twice the period of the signal with peak power was adopted.  Figure \ref{fig:ztf} shows periodograms and phased lightcurves from ZTF photometry of four host stars for which new rotation periods are identified and reported in this work.  Fig. \ref{fig:asas-sn} shows the ASAS-SN data for four stars for which significant ($p < 0.01$) signals with the same period appear in both $g$- and $V$-band photometry.  In the case of GJ 486, the recovered signal at 13.7 days differs markedly from the value of $49.9 \pm 5.5$ days published by \citet{Caballero2022}. 

\bsp	
\label{lastpage}
\end{document}

%% file: newcommands.tex
\newcommand{\halpha}{H$\alpha$}

\newcommand{\tess}{\emph{TESS}}

\newcommand{\gaia}{\emph{Gaia}}
\newcommand{\ktwo}{\emph{K2}}
\newcommand{\jwst}{\emph{JWST}}
\newcommand{\kepler}{\emph{Kepler}}

\newcommand{\msun}{M$_{\odot}$}

\newcommand{\teff}{\ensuremath{T_{\rm eff}}}

